\documentclass[a4paper,11pt]{article}
\pdfoutput=1 % if your are submitting a pdflatex (i.e. if you have
\usepackage{jcappub}

\usepackage{qubic}
\usepackage{xcolor}

\title{QUBIC III: Laboratory Characterization}
\author[1,2]{S.A.~Torchinsky}
\author[1]{J.-Ch.~Hamilton}
\author[1]{M.~Piat}
\author[3,4]{E.S.~Battistelli}
\author[1]{C.~Chapron}
\author[3,4]{G.~D'Alessandro}
\author[3,4]{P.~de~Bernardis}
\author[3,4]{M.~De~Petris}
\author[5,6]{M.M.~Gamboa Lerena}
\author[7]{M.~Gonz\a'{a}lez}
% ...

\affiliation[1]{Universit\'e de Paris, CNRS, Astroparticule et Cosmologie, F-75006 Paris, France}
\affiliation[2]{Observatoire de Paris, Universit\'e Paris Science et Lettres, F-75014 Paris, France}
\affiliation[3]{Universit\a`{a} di Roma - La Sapienza, Roma, Italy}
\affiliation[4]{INFN sezione di Roma, 00185 Roma, Italy}
\affiliation[5]{Facultad de Ciencias Astron\a'{o}micas y Geof\a'{i}sicas (Universidad Nacional de La Plata), Argentina}
\affiliation[6]{CONICET, Argentina}
\affiliation[7]{Centro At\a'{o}mico Bariloche and Instituto Balseiro (CNEA), Argentina}
% ...

\emailAdd{satorchi@apc.in2p3.fr}
\emailAdd{hamilton@apc.in2p3.fr}
\emailAdd{piat@apc.univ-paris7.fr}

\abstract{
We report on an extensive test campaign of a prototype version of the
QUBIC (Q~\&~U Bolometric Interferometer for Cosmology) instrument,
carried out at Astroparticle Physics and Cosmology (APC) in
Paris. Exploiting the novel concept called bolometric interferometry,
QUBIC is designed to measure the CMB polarization at 150~and 220~GHz
from a high altitude site at Alto Chorillo, Argentina. The prototype
model called QUBIC Technological Demonstrator (QUBIC-TD) operates in a
single frequency band (150~GHz) and with a reduced number of
baselines, but it contains all the elements of the QUBIC instrument in
its final configuration. The test campaign included measurements of
the synthesized beam and of the polarization performance, as well as a
verification of the interference fringe pattern. A modulated,
frequency-tunable millimetre-wave source was placed in the telescope
far-field and was used to simulate a point source. The QUBIC-TD field
of view was scanned across the source to produce beam maps. Our
measurements confirm the frequency--dependent behaviour of the beam
profile, which gives QUBIC the possibility to do spectral imaging. The
measured polarization performance indicates a cross-polarization
leakage less than 0.6\%. We also successfully tested the polarization
modulation system, which is provided by a rotating half wave
plate. \nnnew{We demonstrate the full mapmaking pipeline using data
from this measurement campaign, effectively giving an end-to-end
checkout of the entire QUBIC system, including all hardware
subsystems, their interfaces, and the software to operate the whole
system and run the analysis.}  Our results confirm the viability of
bolometric interferometry for measurements of the CMB polarization.}

\date{\today}

\begin{document}

\maketitle

\section{Introduction}
\label{sec:intro}

%\subsection{Scientific Motivation for QUBIC}
\label{sec:motivation}
%% Para 1
The detection of \new{primordial} \bmode\ polarization in the cosmic
microwave background (CMB) is the subject of a worldwide effort due to
its importance as a confirmation of the inflationary model of
cosmology.  A clear detection of polarization \bmode{s} in the CMB at degree angular scales,
\new{distinguished from \bmode{s} incurred by foreground effects \nnew{(including lensing)},} is
evidence of primordial gravitational waves expected during the
inflationary phase in the earliest moments of the universe.  For an
overview see Kamionkowski \& Kovetz~\cite{2016ARA&A..54..227K}.

\nnew{From an observational point of view, the linear polarization of the CMB is described by the Q and U~Stokes parameters than can be transformed mathematically into the scalar E and B-modes respectively corresponding to the curl and curl-free part of the polarization field on the sky. E-modes have been well measured to be of
  order $\sim 1\,\mu$K~\cite{2002Natur.420..772K,2021arXiv210101684D,2020A&A...641E...1A}, as expected from the
  theory, while the \bmode\ component will be an even weaker signal,
  at least an order of magnitude below the \emode~\cite{2021bicep}. As a result, the
measurement of \bmode\ polarization is a difficult exercise of
extracting a signal buried deep within other signals, be them astrophysical foregrounds or instrumental systematics mixing Q and U observables resulting into a leakage of the large E-modes into the small B-modes.}

\label{sec:BIintro}
\nnew{The QUBIC instrument uses a novel technology, called Bolometric Interferometry, that combines the high sensitivity offered by bolometers to the signal purity from interference fringes measurement.
QUBIC has degree-scale angular resolution in order to target the
primordial CMB \bmode\ polarization.  The instrument is designed with
particular attention to the limitation and control of systematic
effects~\cite{2016arXiv160904372A}.  See also in this series of papers
O'Sullivan~et~al.~\cite{2020.QUBIC.PAPER8} for the optics design,
Masi~et~al.~\cite{2020.QUBIC.PAPER5} for the cryogenics design, and
Piat~et~al.~\cite{2020.QUBIC.PAPER4} for the detectors and readout
electronics.  QUBIC uses the technique of interferometry which leads
to the possibility of doing ``self-calibration'', a procedure that
ensures fine control of instrumental systematic
effects~\cite{2013A&A...550A..59B}.  \new{Furthermore, the resulting
  frequency-dependent beam profile on the sky leads to the
  possibility} of doing \new{spectral imaging}~(see
Mousset~et~al.~\cite{2020.QUBIC.PAPER2} in this series of papers).
Bolometric interferometry is the marriage of techniques bringing
together the great sensitivity and large bandwith of bolometers and
the instrumental control and high fidelity imaging of aperture
synthesis.  Using this innovative approach, any residual
\new{systematic effect} in the data will be largely independent from
those in other experiments, thus providing a unique dataset in the
context of the worldwide experimental effort.}

\label{sec:organization}
%% Para 12
\nnew{This article describes the laboratory characterization and calibration phase QUBIC has undergone. }
The paper is organized as follows.
%{Section~\ref{sec:intro} gives an introduction to the cosmological science which motivates the QUBIC experiment.}
Section~\ref{sec:qubicdescription} gives an overview of the
\nnnew{Q~\&~U Bolometric Interferometer for Cosmology (QUBIC)
  Technological Demonstrator} instrument including an introduction to
the concept of bolometric interferometry.  The laboratory setup is
described in section~\ref{sec:calsource} with particular attention
given to the placement and alignment of the calibration source.  This
is followed by section~\ref{sec:results} describing various
measurement results and analysis.  Section~\ref{sec:spectral} presents
the measured bandpass of QUBIC.  Section~\ref{sec:hwp} shows the
measured response to modulated polarization using the rotating half
wave plate (HWP), a critical element to assess the system polarization
performance.
%The QUBIC design provides extremely pure polarization performance with 0.4\%
%cross polarization leakage.
In section~\ref{sec:selfcal} we discuss the measurement of interference fringes on the 
focal plane using the mechanical switches in the horn array to select baselines.
%% and we show an early application of
%% \new{self-calibration} with the independent determination of the
%% physical orientation of the horn array using the fringe measurements.
In section~\ref{sec:syntheticbeam} we show measurements of the QUBIC
synthesized beam, and in section~\ref{sec:mapmaking} we demonstrate
the results of the QUBIC mapmaking pipeline by generating a ``sky''
map of the calibration source using the calibration information
determined from our measurements.  Finally, some concluding
remarks are given in section~\ref{sec:conclusion}.

%% \nnnnew{Throughout this paper, we use our in-house Python-based
%%   software called
%%   \swname{qubicsoft}\footnote{\href{https://github.com/qubicsoft}{https://github.com/qubicsoft}}
%%   for data analysis, and for hardware control and monitoring.  The
%%   \swname{qubicsoft} package is also used for data simulations, except
%%   where otherwise indicated.}

\new{\section{The QUBIC Instrument}
\label{sec:qubicdescription}
%% Para 6
\nnew{QUBIC is an imaging interferometer which measures ``visibilities'' which are the
complex (amplitude and phase) correlations between each antenna pair
(baseline).  In radio astronomy, the visibilities are recorded directly.
A~``correlator'' digitizes the signals and multiplies pairs of signals
to produce a stream of complex numbers, each of which corresponds to
the cross correlation product of an antenna-pair.  Channelization of
the \nnew{spectral} bandpass permits signal processing of individual, very narrow
bands, and for each channel the signal is nearly monochromatic.  In
radio astronomy, large bandwidths are achieved by adding more digital
electronics.}
\subsection{Bolometric Interferometry}

%% Para 7
A bolometric interferometer takes advantage of the high sensitivity
and large bandwidth of bolometers while also benefiting from the
calibration technique possible with an imaging interferometer.
\new{The QUBIC focal \nnew{planes} are populated with transition-edge sensor
  (TES) bolometers (see Piat~et~al.~\cite{2020.QUBIC.PAPER4} in this special
  issue for more details).} The spatial sampling of the sky is
generated by placing a cluster of back-to-back horns that behave
effectively as electromagnetic nozzles.  This horn cluster creates the
$u-v$ sampling of the aperture plane equivalent to what is done by a
distribution of antennas in a radio array.  For the bolometric
interferometer, instead of sampling the signals and computing the
cross correlations between antenna pairs, the interference pattern is
imaged.

%% Para 8
A single image of the interference pattern has all the information
convolved together resulting in observing the sky through a
synthesized beam. The shape of this synthesized beam is given by the
combination of all individual baselines (all pairs of horns).  The
bolometric interferometer ends up being a synthesized imager observing
the sky through its synthesized beam just the same way as a classical
imager observes the sky through the beam formed by the telescope.  For
calibration and instrumental systematic effects studies it is
however crucial to extract the individual visibilities.  By blocking
all horns except two, we would measure the interference pattern
of that baseline.  For technical reasons in the QUBIC Technological
Demonstrator (QUBIC-TD), only two horn shutters can be shut at a time,
but by making a series of measurements with different pairs of horns
blocked, the result is equivalent to having all horns blocked except
two~\cite{2013A&A...550A..59B}.  The $20\times20$ cluster of horns in
the QUBIC Full Instrument (QUBIC-FI) has 400~horns making
\mbox{$n(n-1)/2 = 79800$} baselines which are needed to be observed
individually for self-calibration.  The QUBIC-TD, whose tests are
reported in this paper, has a smaller horn array with 64~horns in an
$8\times8$ square array giving 2016~baselines.

%% Para 9
\new{The main parameters which determine the general performance of
  the bolometric interferometer are the number of baselines and the
  number \nnnew{of} detectors.}  Performance of the bolometric interferometer
improves as the number of baselines increases since this improves the
application of self-calibration by increasing the number of
parameters.  It also leads to a more sharply defined synthesized beam.
A larger cluster of horns provides more baselines, but this in turn
must be sampled by a larger array of detectors in the focal plane.
\nnew{A larger horn cluster is also effectively a larger telescope
  aperture which increases the power throughput of the system.
  Increasing the number of detectors therefore not only improves the
  sampling of focal plane but also improves the sensitivity of the
  instrument by detecting more power.}  As a result, the overall
\new{performance} of the bolometric interferometer is \new{not simply}
a function of the number of detectors in the focal plane.  \new{It
  also depends on the number of baselines formed by the horn array
  which must be large enough to match the surface covered by detectors
  in the focal plane.}

%% Para 10
An additional feature of bolometric interferometry is the possibility
to do \new{spectral imaging}.  The synthesized beam is predicted to
vary in a well-understood way with frequency.  In particular, the beam
secondary lobes \new{are closer to the central lobe for higher
  frequency of incident radiation} while the central lobe remains in
the same place \new{across the spectral band}.  As a result, the time
ordered data (TOD) effectively samples different \new{electromagnetic}
frequencies as the beam passes over the same point in the sky.  \new{An
important objective of our calibration campaign was to verify this
behaviour with the QUBIC-TD.}

%% Para 11
This \new{spectral} selectivity can be deconvolved
in the data post processing.  The spectral resolution improves with
the number of baselines as the synthesized beam has finer secondary
lobes with a larger cluster of horns.  Spectral imaging is an
innovative feature of bolometric interferometry which gives QUBIC an
important advantage over other CMB imagers (see
Hamilton~et~al.~\cite{2020.QUBIC.PAPER1} and Mousset~et~al.~\cite{2020.QUBIC.PAPER2} for details) and \new{the
  expected frequency dependence of QUBIC's synthesized beam} has been
confirmed (see section~\ref{sec:syntheticbeam}).

\subsection{Design Overview}
%% Para 1
QUBIC employs an optical system consisting of back-to-back horns that
select the relevant baselines and an optical combiner focusing on a
bolometric focal plane. The optical combiner forms interference
fringes while the bolometers average their powers over timescales much
larger than the period of the light waves. This is therefore the
optical equivalent of a wide-band correlator in classical
interferometry. The instrument operates at cryogenic temperatures
thanks to a large cryostat described in
Masi~et~al.~\cite{2020.QUBIC.PAPER5}.

\begin{figure}[t]
\centering
\includegraphics[width=0.95\linewidth]{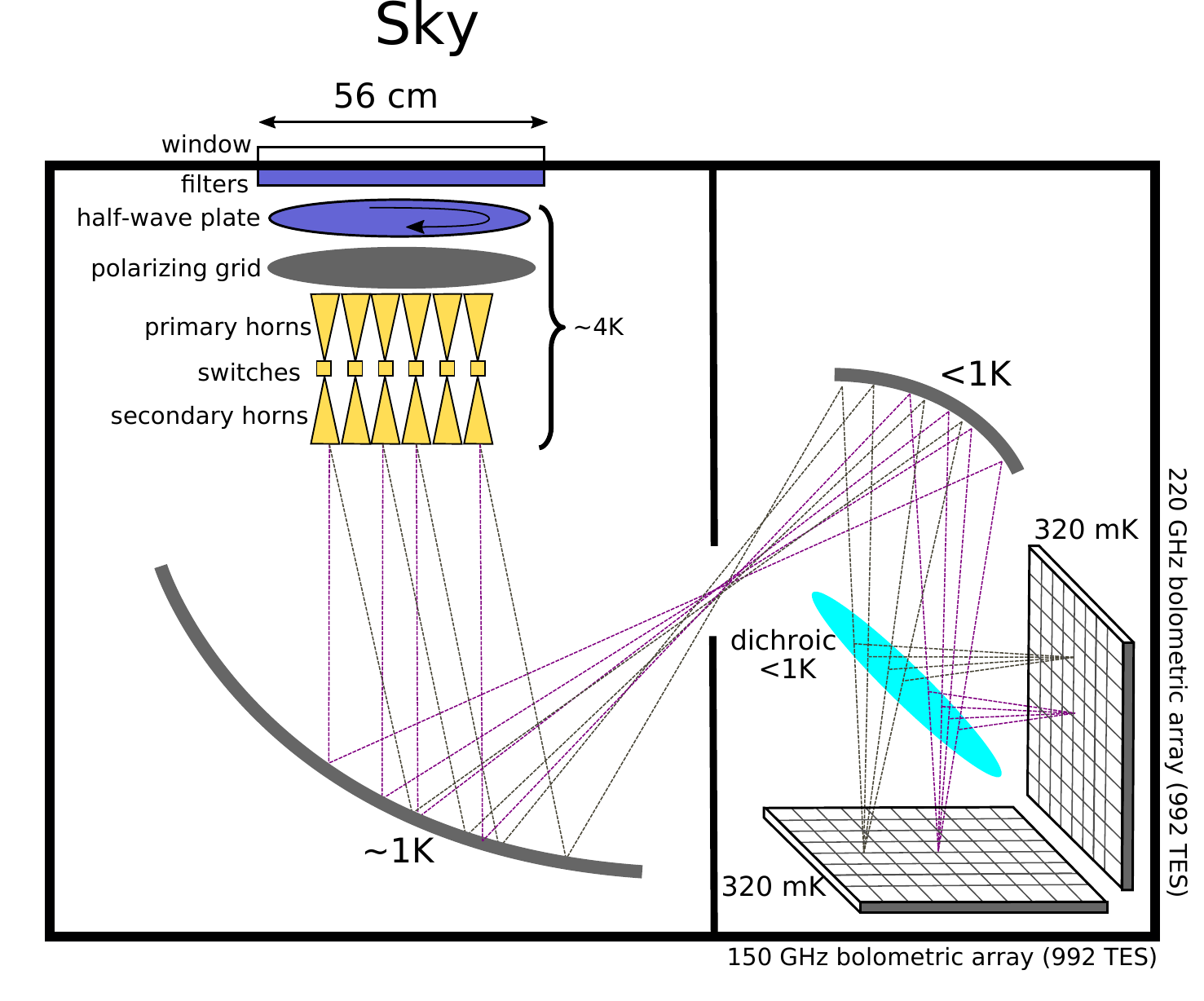}\\
\includegraphics[width=0.95\linewidth]{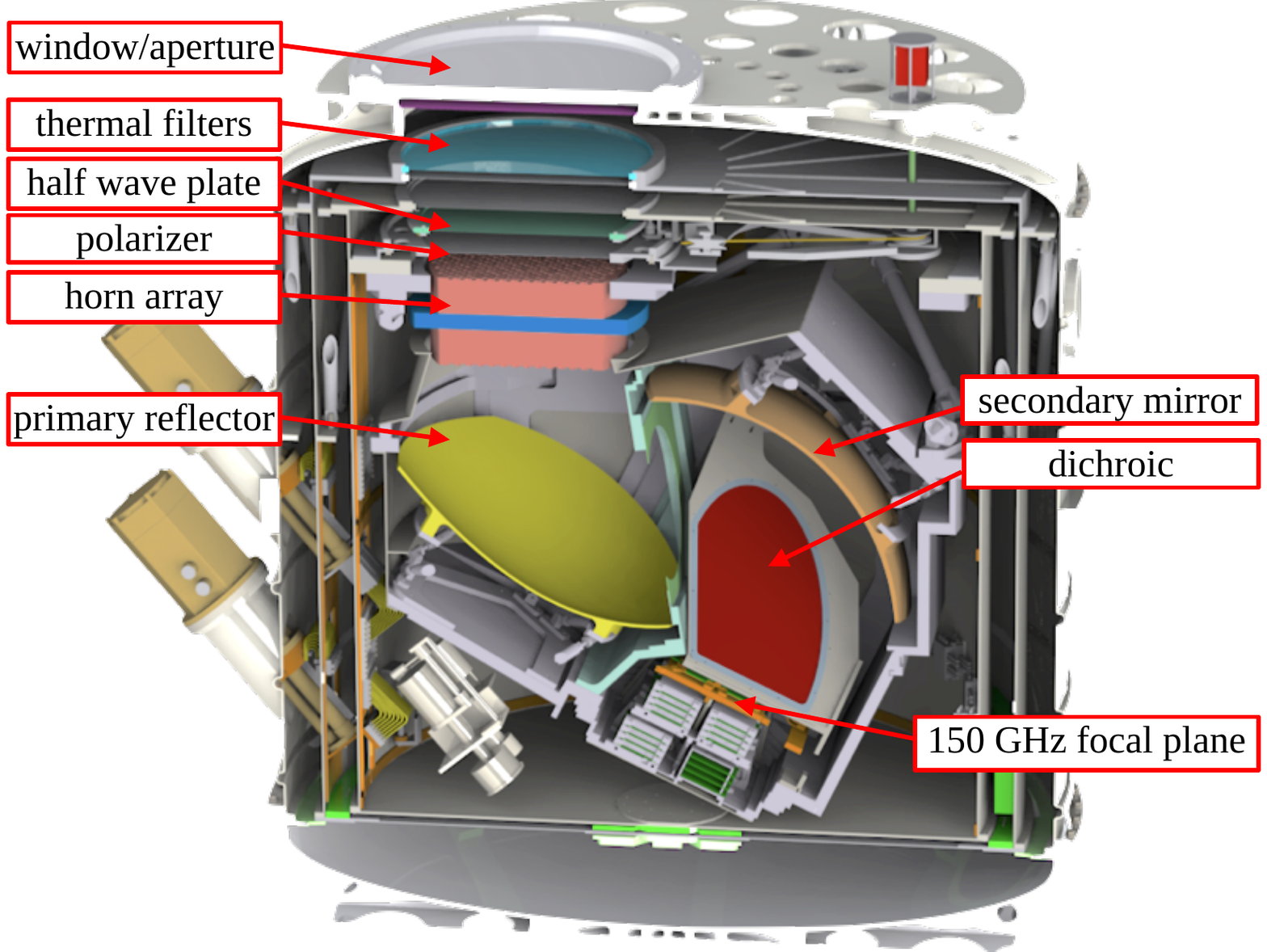}
\caption{\new{Schematic of the QUBIC\nnew{-FI} instrument (above) and sectional cut of
  the cryostat (below) showing the same sub-systems in their real
  configuration.  \nnew{The second detector array is not shown in the CAD rendered image below.}}\label{fig:qubicschematic}}
\end{figure}

\newlength{\colone}
\settowidth{\colone}{Synthesized beam FWHM [degrees]~~~}
\newlength{\coltwo}
\settowidth{\coltwo}{[131-169]~GHz}
\newlength{\colthree}
\settowidth{\colthree}{0.39 (150~GHz), 0.27 (220~GHz)}

\begin{table}[t]
    \renewcommand{\arraystretch}{1.}
    \begin{center}
        \caption{\label{table:qubic_params}\new{QUBIC main parameters}}
\begin{tabular}{p{\colone} p{\coltwo} p{5.5cm}}
          \hline
            Parameter& {\qtd} & {\qfi} \\
            \hline
            \hline
            Frequency channels \dotfill &150 GHz & 150 GHz \& 220 GHz\\
            Frequency range 150 GHz \dotfill &[131-169] GHz &[131-169] GHz\\
            Frequency range 220 GHz \dotfill &- &[192.5-247.5] GHz\\
            Window Aperture [m]\dotfill & 0.56 & 0.56 \\
            Number of horns\dotfill &64 &400\\
            Number of detectors\dotfill &248 &992$\times$2\\
            Detector noise [$\mathrm{W/\sqrt{Hz}}$]\dotfill & 2.05$\times 10^{-16}$ & 4.7$\times 10^{-17}$ \\
            Focal plane temp. [mK]\dotfill &300 &300\\
            Sky Coverage\dotfill &1.5\% &1.5\%\\
            Synthesized beam FWHM [degrees]\dotfill &0.68 &0.39 (150~GHz), 0.27 (220~GHz)\\
        \end{tabular}
        
    \end{center}
\end{table}

%% Para 2
A schematic of the design of \nnew{the {\qfi}} is shown in
figure~\ref{fig:qubicschematic} and the main instrument parameters are
listed in Table~\ref{table:qubic_params}. \nnew{The {\qtd} has the
  same optical layout and differs only in the number of horns and
  detectors, and it lacks the dichroic and second focal plane for
  220~GHz.} The sky signal first goes through a 56~cm diameter window
made of ultra high molecular weight polyethylene followed by a series
of \nnnew{low-pass} filters cutting off frequencies outside the desired band.
\nnew{The window is not anti-reflection coated.}  \nnew{The filters
  are Cardiff metal mesh filters (see O'Sullivan
  et~al.\cite{2020.QUBIC.PAPER8} for details).}

%% Para 3
The next optical component is the stepped rotating HWP which modulates
incoming polarization (see~D'Alessandro~et~al.~\cite{2020.QUBIC.PAPER6} for details).
\nnnew{The HWP is a metal-mesh device as described by Pisano~et~al.~\cite{pisano_hwp_meta}.}
A
single linear polarization is selected just after polarization
modulation by a wire-grid.  In this scheme, any unwanted
cross-polarization which might be introduced by components later in
the optical chain do not affect the result since the bolometers absorb
all radiation, independent of polarization.

%% Para 4
The next optical device \nnew{in the {\qfi}} is an array of
400~back-to-back corrugated horns made of an assembly of two 400-horn
arrays, composed of 175~aluminium platelets (0.3~mm thick) chemically
etched to reproduce the corrugations required for the horns to achieve
the required performance. \nnew{The horn mouth diameter is 12~mm and
  the spacing between horn centres on a row is 14~mm.  For the {\qtd},
  there are only 64~horns.}  Both front and back horns are identical
with a field of view of 13~degrees FWHM with secondary lobes below
$-$25~dB~\cite{2020.QUBIC.PAPER7}.  An array
of mechanical shutters (RF switches) separates the two back-to-back
horn arrays in order to be able to close or open horns for
self-calibration (see section~\ref{sec:selfcal}). The shutters are
spring loaded and activated by applying a voltage to an induction
coil.  As a result, the shutter requires continuous electrical current
in order to remain closed.  This dissipates some heat such that it is
only possible to activate a maximum of two shutters at a time without
over heating the horn array which is maintained at 1~K.

%% Para 5
The back-horns directly illuminate the two-mirror off-axis Gregorian
optical combiner~\cite{2020.QUBIC.PAPER8}
which focuses the signal onto the two perpendicular focal planes
\nnew{in the {\qfi} and only one focal plane in the {\qtd}.  In the
  {\qfi},} a dichroic filter splits the incoming waves into two wide
bands centred at 150~GHz for the on-axis focal plane and 220~GHz for
the off-axis focal plane. The \nnew{{\qfi}} focal planes are each
equipped with 992~NbSi TES~\cite{2020.QUBIC.PAPER4} cooled to 320~mK using a
\new{double-stage $^3\mathrm{He}/^4\mathrm{He}$ sorption cooler~\cite{2020.QUBIC.PAPER5}.  The
    {\qtd} has only 248~TES.}

%% Para 6
\nnnew{A schematic view and CAD model} of the cryostat can be seen in
figure~\ref{fig:qubicschematic}. The cryostat weighs roughly 800~kg
and is around 1.6~m high with a 1.4~m diameter.  The {\qtd} includes
all the critical elements of the {\qfi}, thus making the test campaign
presented here representative of the complete configuration. The
{\qtd} uses the same cryostat, cooling system, filters and general
sub-system architecture as the {\qfi} but with only 64~back-to-back
horns and smaller mirrors to match the illumination of the
\mbox{$8\times8$~horn-array}. It has a single 248~TES bolometer array
operating at 150~GHz.  \nnew{The \mbox{{\qfi}} will have improved
  performance because of the increased size of the horn cluster and
  the increased number of detectors, as explained in
  section~\ref{sec:BIintro}.  Also, the {\qfi} detectors will be
  background-limited.  The {\qtd} detection chain is currently not
  background-limited because of aliasing in the readout
  system~\cite{2020.QUBIC.PAPER4}.  The {\qfi} will have increased
  sampling frequency and will add Nyquist inductors to mitigate this
  effect.}

\nnnnew{Throughout this paper, we use our in-house Python-based
  software called
  \swname{qubicsoft}\footnote{\href{https://github.com/qubicsoft}{https://github.com/qubicsoft}}
  for data analysis, and for hardware control and monitoring.  The
  \swname{qubicsoft} package is also used for data simulations, except
  where otherwise indicated.}
}

\clearpage
\section{Calibration Source}
\label{sec:calsource}
\subsection{Setup for QUBIC-TD}
%% Para 1
Characterization of the QUBIC-TD instrument is done primarily using a
frequency-tunable monochromatic point source in the far-field.  This
permits measurement of the bandpass (section~\ref{sec:spectral}), the
polarization performance (section~\ref{sec:hwp}), the measurement of
interference fringes (section~\ref{sec:selfcal}), and the beam point spread function
(PSF, section~\ref{sec:syntheticbeam} and
section~\ref{sec:mapmaking}).  This section describes the optical
setup used to characterize QUBIC-TD.

\begin{figure}[t]
  \centering
  \nnewfig{\includegraphics[width=0.95\linewidth]{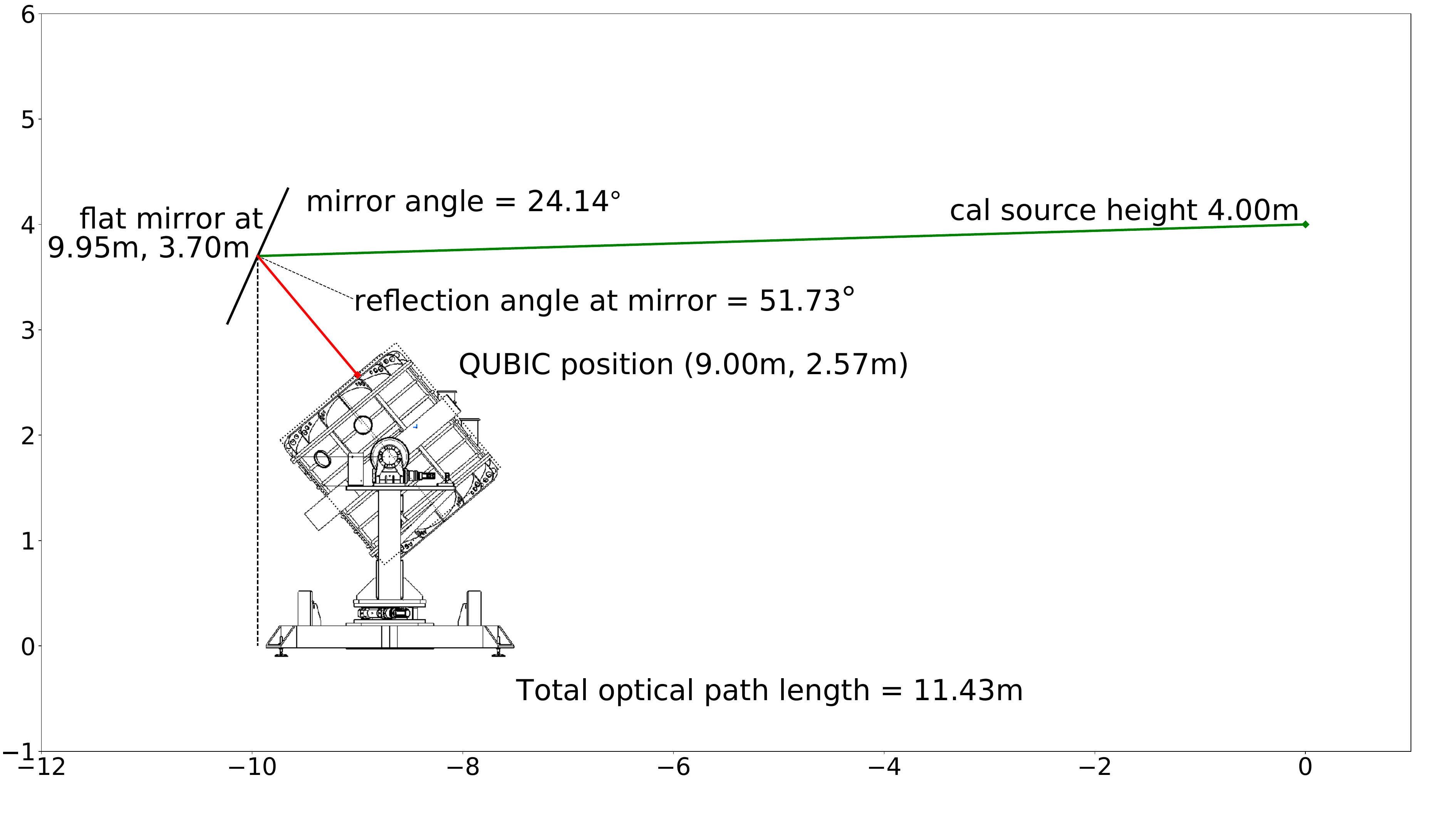}}
  \caption{Layout for the calibration source relative to the QUBIC instrument.\label{fig:calsource_sketch}}
\end{figure}

%% Para 2
The setup is shown in the sketch of figure~\ref{fig:calsource_sketch}.
The calibration source points at a flat mirror which redirects the
beam into the QUBIC cryostat window.  There is an 11~m optical path
putting the calibration source effectively in the far field (see
section~\ref{sec:farfield}).

\begin{figure}[t]
  \centering
  \begin{minipage}{0.49\linewidth}
    \flushleft
    \includegraphics[width=0.95\linewidth]{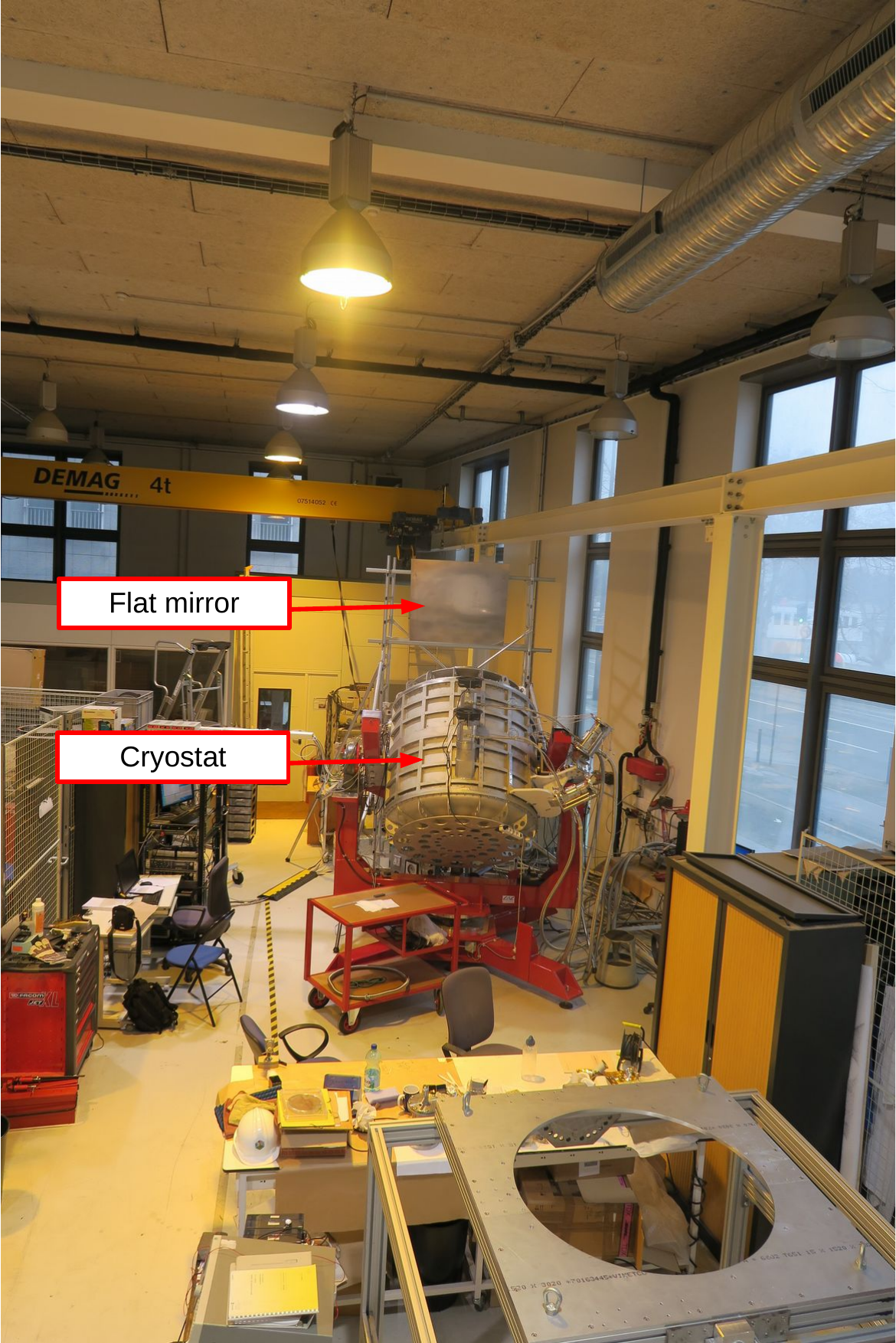}
  \end{minipage}
  \begin{minipage}{0.49\linewidth}
    \flushright
    \includegraphics[width=0.95\linewidth]{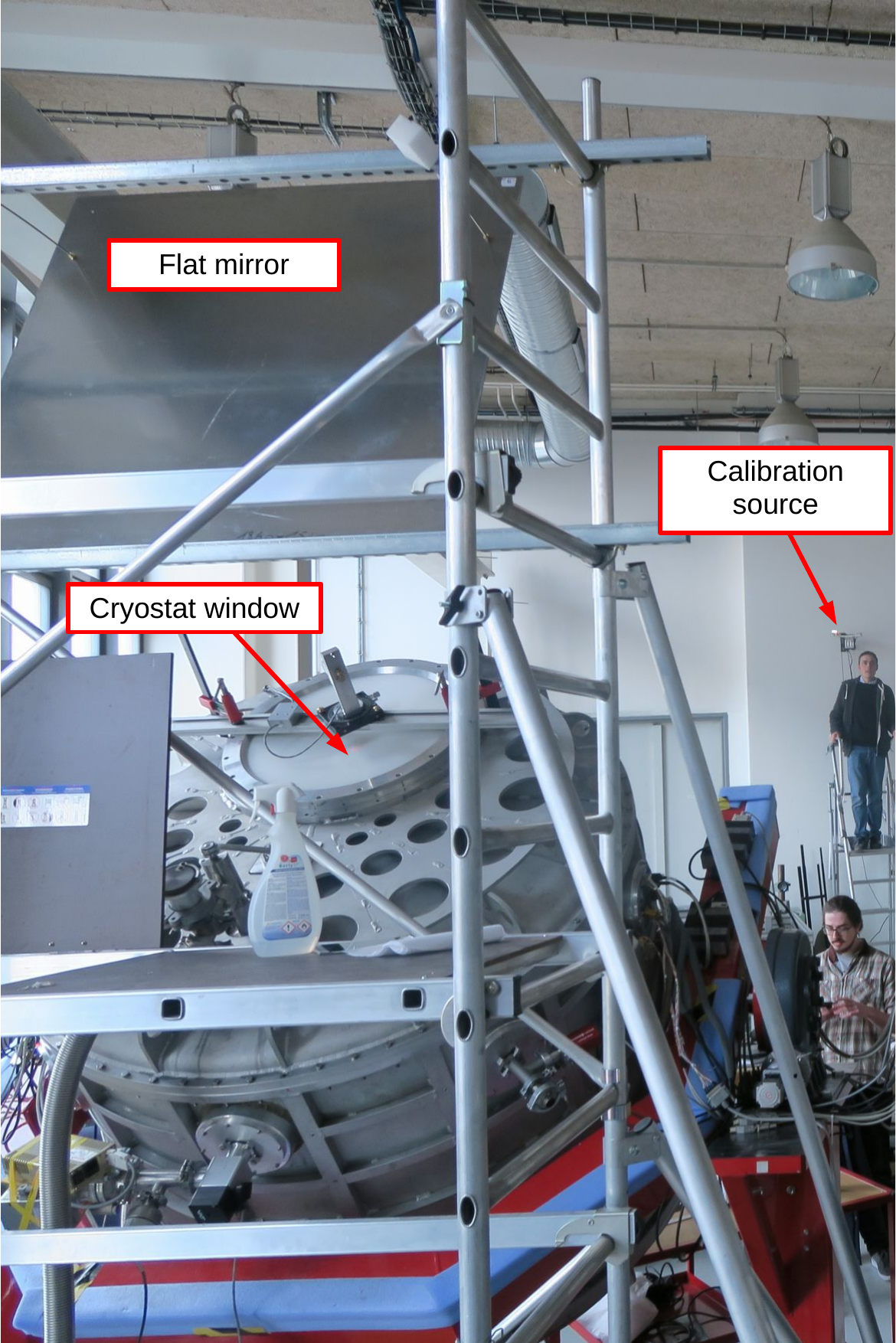}
  \end{minipage}
  \caption{
    {\textbf{left:}~~Photo of QUBIC looking along the line-of-sight from the
      calibration source.  The reflection of the window is clearly
      visible in the flat mirror.\label{fig:calsource_view_of_qubic}}
    {\textbf{right:}~~The flat mirror is mounted on a scaffold at a height of
      3.7~m and has a finely adjustable tilt angle using a system
      composed of a long screw.\label{fig:calsource_alignment2}}
  }
\end{figure}

%% Para 3
The flat mirror is an aluminium sheet mounted on a scaffolding at a height
of 3.7~m (photo in figure~\ref{fig:calsource_alignment2}).  The tilt angle of
the flat mirror can be adjusted by a long screw ensuring
precise selection for the correct tilt angle.

\begin{figure}[t]
  \centering
  \newfig{\includegraphics[width=0.95\linewidth]{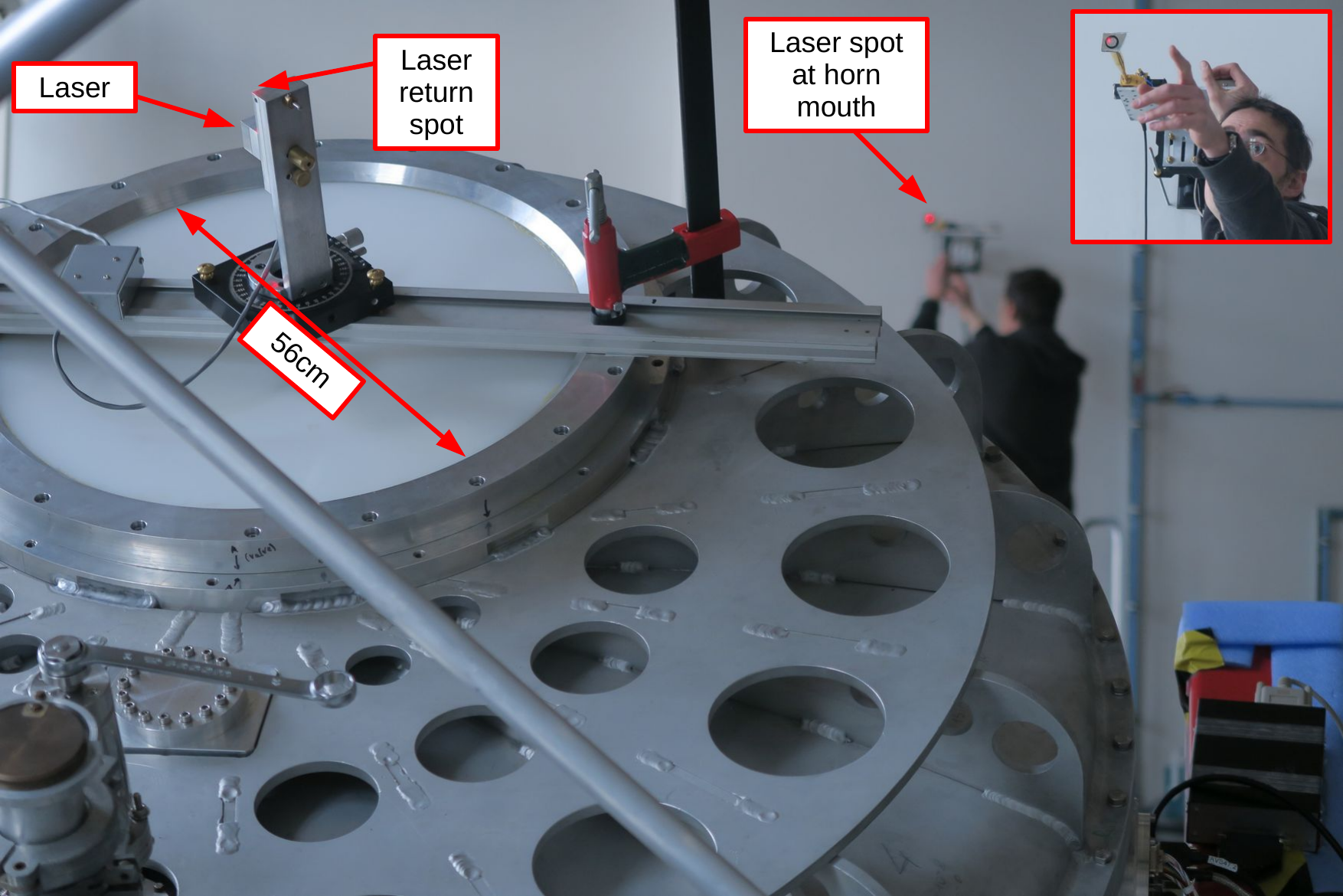}}
  \caption{ {Photo of the alignment procedure.  The laser is mounted
      orthogonal to the cryostat window and shines towards the flat
      mirror, sending the light to the calibration source feedhorn at
      the other side of the room.  A small mirror fitted to the mouth
      of the feedhorn sends the laser light back on the same path to
      the cryostat window.  The laser spot is clearly visible on the
      laser mount structure.  \new{The inset, top-right, shows a
        close-up of the calibration source with a mirror mounted on
        the horn mouth and the laser spot within the area of the horn
        mouth.}\label{fig:calsource_alignment1}\label{fig:laser_on_horn}}
  }
\end{figure}

%% Para 4
Alignment of the system was accomplished using a laser temporarily
mounted at the window of the QUBIC cryostat, pointing normal to the
window (photo figure~\ref{fig:calsource_alignment1}).  The laser light
is reflected from the flat mirror to the calibration source where a
small flat mirror was fitted to the front of the calibration source
feedhorn.  The laser reflects from the mirror at the feedhorn mouth
and returns to the large flat mirror and finally to the QUBIC cryostat
window.  In the photo (figure~\ref{fig:calsource_alignment1}), one can
clearly see the spot of the laser on the corner of the laser mounting
structure on the window.  The alignment is therefore precise to within
a fraction of a degree which is well within the tolerance necessary to
have the calibration source visible to the QUBIC-TD.

\begin{figure}[t]
  \centering
  \includegraphics[height=0.45\textheight]{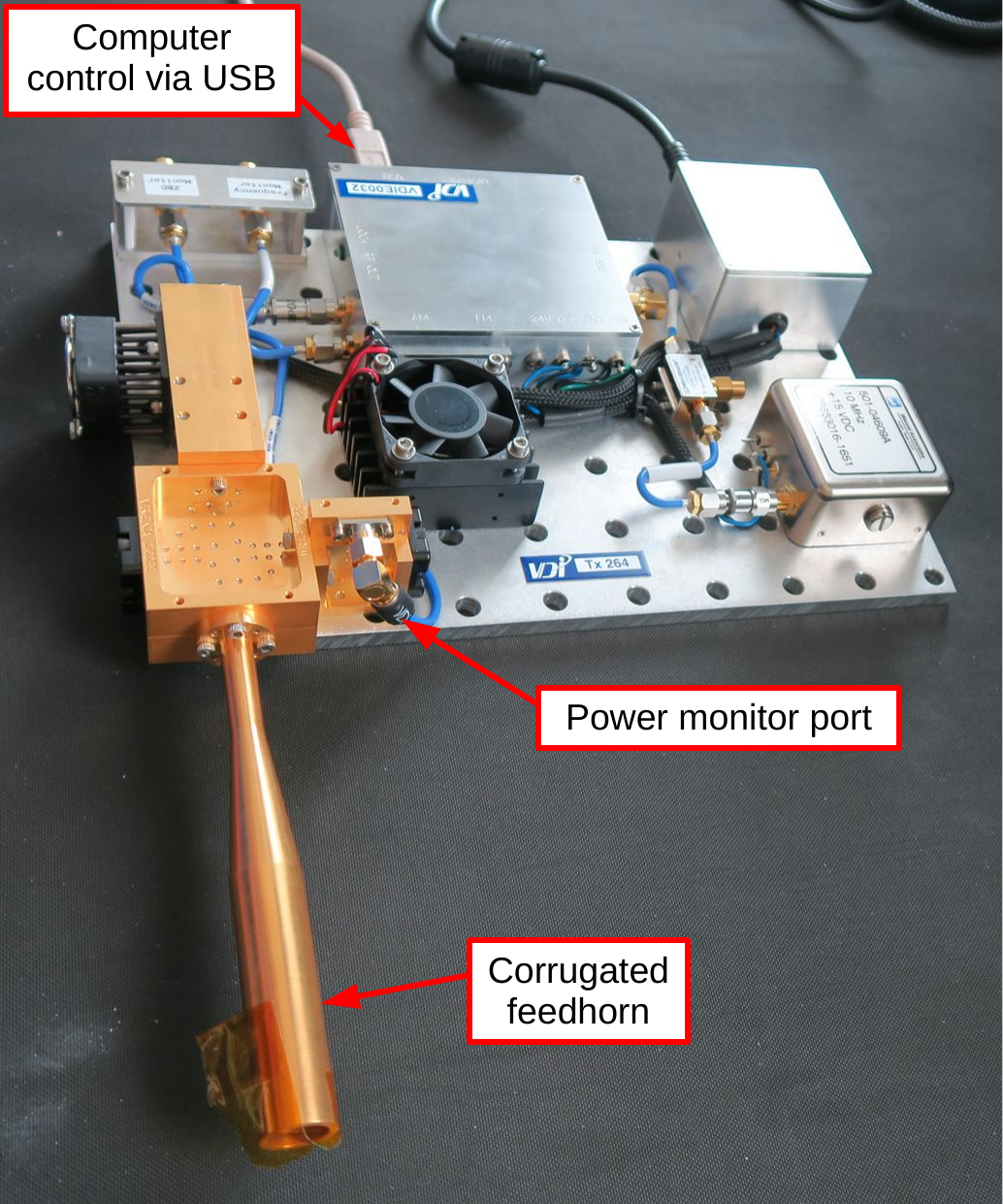}
  \caption{{Photo of the VDI millimetre-wave source.  The electronics}
    include a signal generator for modulating the calibration source
    with a square wave. The output power \new{monitor} of the
    calibration source is sent through an amplifier and then digitized
    by an ADC board integrated in the Raspberry~Pi (seen suspended on
    the cables below the calibration
    source).\label{fig:calsource_electronics}}
\end{figure}

%% Para 5
The calibration source system is composed by the source itself together with a dedicated electronic subsystem supporting its operations.  The calibration source was purchased
from Virginia Diodes Incorporated (VDI) electronics\footnote{\url{https://vadiodes.com}} and is
composed of a synthesizer providing frequencies around 10~GHz
\nnew{followed by a two-stage multiplication chain}.  \new{A coaxial
  cable from the synthesizer connected by standard SMA connectors
  leads to the first of} two multipliers \new{which are} in a
\new{waveguide-coupled} chain.  \new{Together, they} multiply the base
frequency by~12 resulting in frequencies around 150~GHz.  \new{The
  signal is transmitted by rectangular waveguide after the final
  multiplier to a profiled corrugated feedhorn.}  The range of the
system is between 130~GHz and 170~GHz with a frequency tuning
  resolution of \nnnew{144~Hz}.  The nominal output power is $+9$~dBm and it
  can be modulated with a maximum amplitude of $1.25$~dB. The
  frequency is configured via USB connection by a nearby Raspberry~Pi
mini computer (see photo figure~\ref{fig:calsource_electronics}).

%% Para 6
A signal generator provides a square wave at around 1~Hz which is used
to modulate \new{the amplitude of the} calibration source.  The
Raspberry~Pi configures the signal generator, selecting the amplitude,
offset, frequency, shape, and duty cycle.
%% Para 7
A directional coupler at the waveguide before the entrance to the
corrugated feedhorn provides a port for monitoring the output power of
the calibration source.  \nnnew{The horn return loss is less than
  $-30$~dB, as reported by the supplier of the horn, Custom Microwaves
  Inc.,\footnote{\url{https://custommicrowave.com}} and therefore the monitoring of the output power variation at
  the end of the multiplier chain is a reliable method for monitoring
  the power variation in the calibration source output power
  illuminating the aperture of QUBIC-TD.} A probe on this port
provides a voltage which is a function of the output power.  This
voltage is sent to an analogue-to-digital converter (ADC) board
integrated in the Raspberry~Pi mini computer which \new{broadcasts the
  data on the internel network} along with a timestamp for each
sample.  There are approximately 300~samples per second.
%% Para 8
A command line interface written in Python is used to configure the
calibration source setup.  This system accepts commands via \new{network} socket and
can be easily interfaced by the graphical user interface called
\swname{QubicStudio}.

%%%%%%%%%%% referee comment
%% please provide a minimal set of descriptions (e.g. one paragraph
%% and cite your accompany paper) about the detector condition in the
%% cryostat. Given a source being a high power coherent source, any
%% extra treatment of a detector saturation or bias setting
%% specifically for this lab measurement? And/or does this lab
%% condition provide too high loading, and therefore you need any
%% (extra) IR filter, e.g. neutral density filter, or two step TES
%% transition to accept a higher loading condition? If you use any
%% extra "feature" to work in the lab loading condition, how does this
%% lab demonstration relate to the on-sky performance?

%% Para 9
\new{There is an important difference in the background loading for
  the QUBIC-TD in the lab compared to an on-sky measurement.  The lab
  background loading is of the order of 300~K, while a measurement on
  the sky will have background loading of the order of a few tens
  of~K. In order to reduce the laboratory background loading, the
  QUBIC-TD is equipped with a neutral density filter (NDF) mounted at
  the 1~K stage.  The NDF transmits 9\% of all incoming radiation.
  This reduces the 300~K ambient temperature background to about 27~K
  which approximates what is expected from the sky at the 5000~m above
  sea-level observing site.  As a result, the loading on the
  bolometers in the laboratory measurements with the NDF approximates
  what it will be at the observing site without the NDF.}

%% Para 10
\new{A second difference between the setup in the lab and the planned
  setup for the instrument at its observing site is the power received
  by the calibration source.  The \nnew{VDI} millimetre-wave
  source emits significant power.  In the laboratory setup, it is
  placed at a distance of 11.4~m, while at the observing site, the
  calibration source will be mounted on a tower 40~m distant from the
  base of the QUBIC telescope, giving an optical path length of
  approximately 60~m.  This means that the calibration source detected
  in the lab is roughly 30~times more powerful than in the setup
  planned at the observing site.  \nnnew{For this reason,} Eccosorb material \nnew{AN72} is placed at the
  calibration source feedhorn mouth in order to attenuate the signal
  to approximately 3\% \nnnew{across the band} (see Schillaci~et~al.~\cite{2013InPhT..58...64S} for a
  measurement of attenuation by Eccosorb).}

\begin{figure}[t]
  \centering
  \nnewfig{\includegraphics[width=0.9\linewidth]{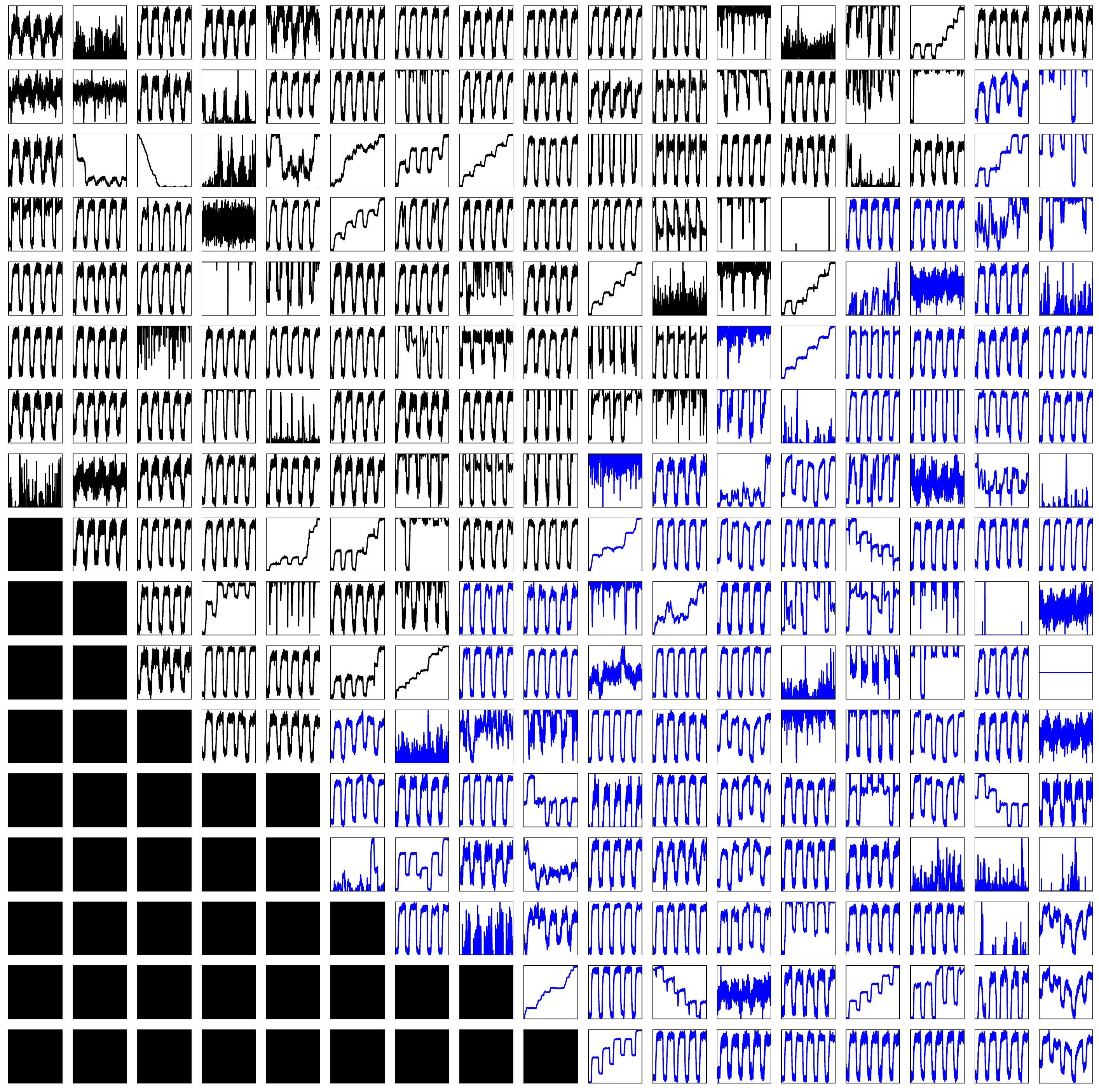}}
  \caption{The first measurement of the calibration source by QUBIC.
    \new{Each box in the plot shows a 20~second timeline for the
      detector in that position in the focal plane.  The vertical axis
      for each plot is in arbitrary power units, scaled for the
      minimum and maximum of each plot.}  The signal is clearly seen
    in most pixels, corresponding to the good pixels in the array (see
    Piat~et~al.~\cite{2020.QUBIC.PAPER4} for a discussion on the
    bolometer array performance).  \new{The curves in blue and black
      differentiate the two read-out electronics chains (128~detectors
      per read-out electronics box).}  The black, \new{filled-in}
    ``pixels'' in the bottom-left are empty positions.  The QUBIC-FI
    will have four arrays equivalent to this one in order to make a
    roughly circular focal plane for each frequency
    channel. \label{fig:First_Cal_Source_Detection}}
\end{figure}

\begin{figure}[t]
  \centering
  \newfig{\includegraphics[width=0.9\linewidth]{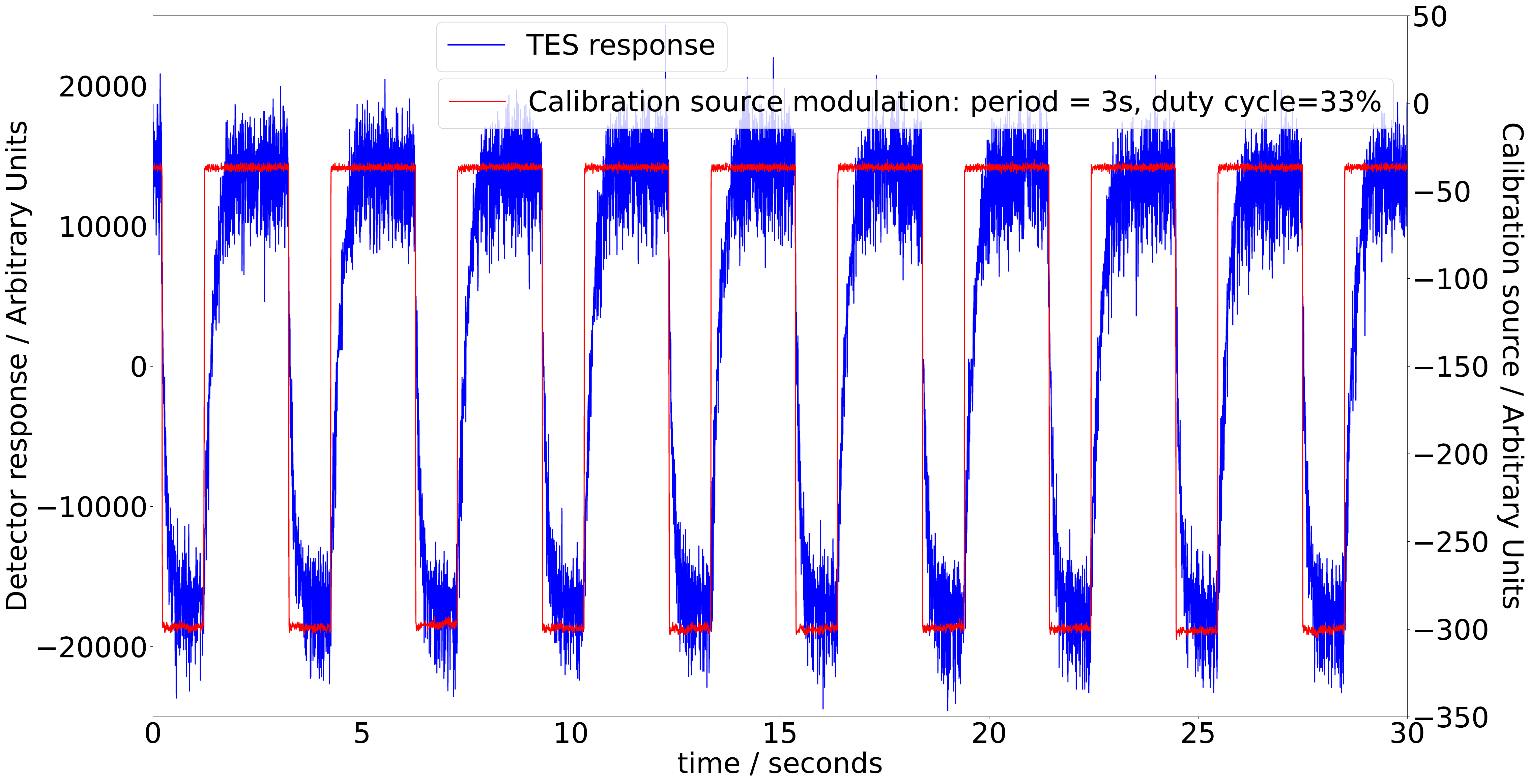}}
  \caption{Overlay of the signal detected by a TES detector of the QUBIC 
    array (blue curve) together with the modulated calibration source signal as
    measured by the calibration source power monitor (red curve).\label{fig:calsource_modulation}}
\end{figure}

%% Para 11
Figure~\ref{fig:First_Cal_Source_Detection} shows the signal response
of all the TES detectors in the QUBIC-TD array.  The source was set at
150~GHz and was modulated with a square wave with 3~second period
(0.333~Hz) and a 33\% duty cycle.  \new{The period was chosen to be
  significantly longer than the expected time constant of the TES
  bolometers, and with a non-symmetrical duty cycle
  (${1}/{3},{2}/{3}$) in order to easily distinguish the OFF and ON
  cycles.}  Figure~\ref{fig:calsource_modulation} shows in detail the signal
measured by a typical well-behaved TES together with the modulation
signal measured by the power monitor.  There is a clear correlation
between the two \new{and the time constant can be estimated to be of
  the order of $40$~ms.  A detailed analysis of the QUBIC-TD bolometer response is given in
  \nnew{Piat~et~al.}~\cite{2020.QUBIC.PAPER4}.}

%%%%%%%% referee comment
%% Section 7, I suspect that the lab beam measurements using the
%% coherent source are susceptible to the standing wave. Did you
%% concern this effect? Did you mitigate by hardware or software about
%% this effect? If yes, I expect to appear somewhere in this paper.
%%
%% Section 7, did you concern about the scattering of the power from
%% the high signal-to-noise source off the wall and cryostat
%% window/rim?

%% Para 12
While the APC laboratory is not equipped as an anechoic environment
\nnnnew{(see figure~\ref{fig:calsource_view_of_qubic}),}
our measurements show no evidence of reflection effects from the
surroundings.  The mapping of the beam did not show unwanted artefacts
(see, e.g., figure~\ref{fig:synthbeam} in
section~\ref{sec:syntheticbeam}, and also the result of the mapmaking
algorithm, figure~\ref{fig:mapmaking}, in
section~\ref{sec:mapmaking}).  Nevertheless, it is possible that
low-level reflections may have contributed some extra noise to the
image.  \nnnew{The diffuse features around the peaks may be due to
  reflections in the lab, or they are possibly due to instrumental
  imperfections that can be dealt with at the map-making stage once
  they are measured.  These maps are expected to be improved when the
  measurements are made at the observing site where external
  reflections will be at a much lower level, while instrumental
  artefacts remain the same.}

%% Para 13
\new{Another possibility for reflections might come from the
  face-to-face setup of the calibration source and the QUBIC window.
  This could lead to standing-waves but a number of parameters
  mitigate this possibility.  The long optical path between the
  cryostat window and the calibration source imposes significant
  attenuation of the signal after reflection, and multiple reflections
  will have insignificant power.  The mouth of the feedhorn at the
  calibration source subtends an angle of approximately 0.05~degrees,
  therefore exposing negligible area to the cryostat window for
  reflections.  Finally, a layer of Eccosorb \nnew{AN72} was placed in
  front of the calibration source feedhorn mouth for further
  attenuation.  \nnnew{If we use purely geometric optics reasoning,
    which gives an upper limit to the magnitude of standing waves, we
    can simply compare the area of the two face-to-face surfaces.  The
    QUBIC-TD window has a diameter of 56~cm and the calibration source
    feedhorn has a diameter of 1~cm.  The ratio of the areas of these
    surfaces is less than 0.03\%.  Therefore the power is attenuated
    by 0.03\%, without considering the attenuation by the Eccosorb and
    by the distance between the two surfaces.  This is below
    our measurement precision.} For example, the bandpass measurement
  of figure~\ref{fig:bandpass} in section~\ref{sec:spectral} shows no
  evidence of standing waves.}

% text from David Burke explaining that the source is effectively in the far field
\clearpage
\new{% this is text provided by David Burke from his thesis
\subsection{Optical Modelling of the Calibration Source as viewed by QUBIC}
\label{sec:farfield}

%%% new text from Creidhe
\nnnnew{For an accurate analysis of the instrument performance, it is
  important to have a simulation of the expected pattern on the focal
  plane. For this purpose, we use optical modelling with the source
  set at the same distance (11.4~m from the QUBIC-TD window) as in our
  experimental setup.  \nnnnnew{The comparison between measurement and
    model is done regardless of the distance to the calibration
    source,} however it is also useful to have the calibration source
  sufficiently in the far-field so that it appears as an unresolved
  point source resulting in the focal plane pattern being the PSF of
  the instrument.  The PSF has several sharp features with which to
  compare. Once in the far-field the focal plane pattern does not vary
  significantly with the distance to the calibrator.}

Using the software package \swname{MODAL} (Maynooth University, see
O'Sullivan~et~al.~\cite{2020.QUBIC.PAPER8} for details about optics
modelling), a model of the 150~GHz source was positioned 1~m and
10~m from the aperture.  For each position the following quantities
were determined:
\begin{itemize}
\item the incident radiation over the horn array.
\item the horn output (at a distance of 60~cm) for two arbitrarily chosen horns in the $8\times8$ horn array.
\item the resulting PSF on the focal plane formed from the radiation for all 64~horns in the array.
\end{itemize}

Both the amplitude and phase of the incident field from the calibrator
source were investigated at the input of the horn array. It can be
seen in figure~\ref{fig:burke410} \nnnew{that at a distance of
  10~m, the amplitude and phase of the calibrator beam is
  quite uniform over the horn array.}
\begin{figure}[t]
  \centering
  \includegraphics[width=0.95\linewidth]{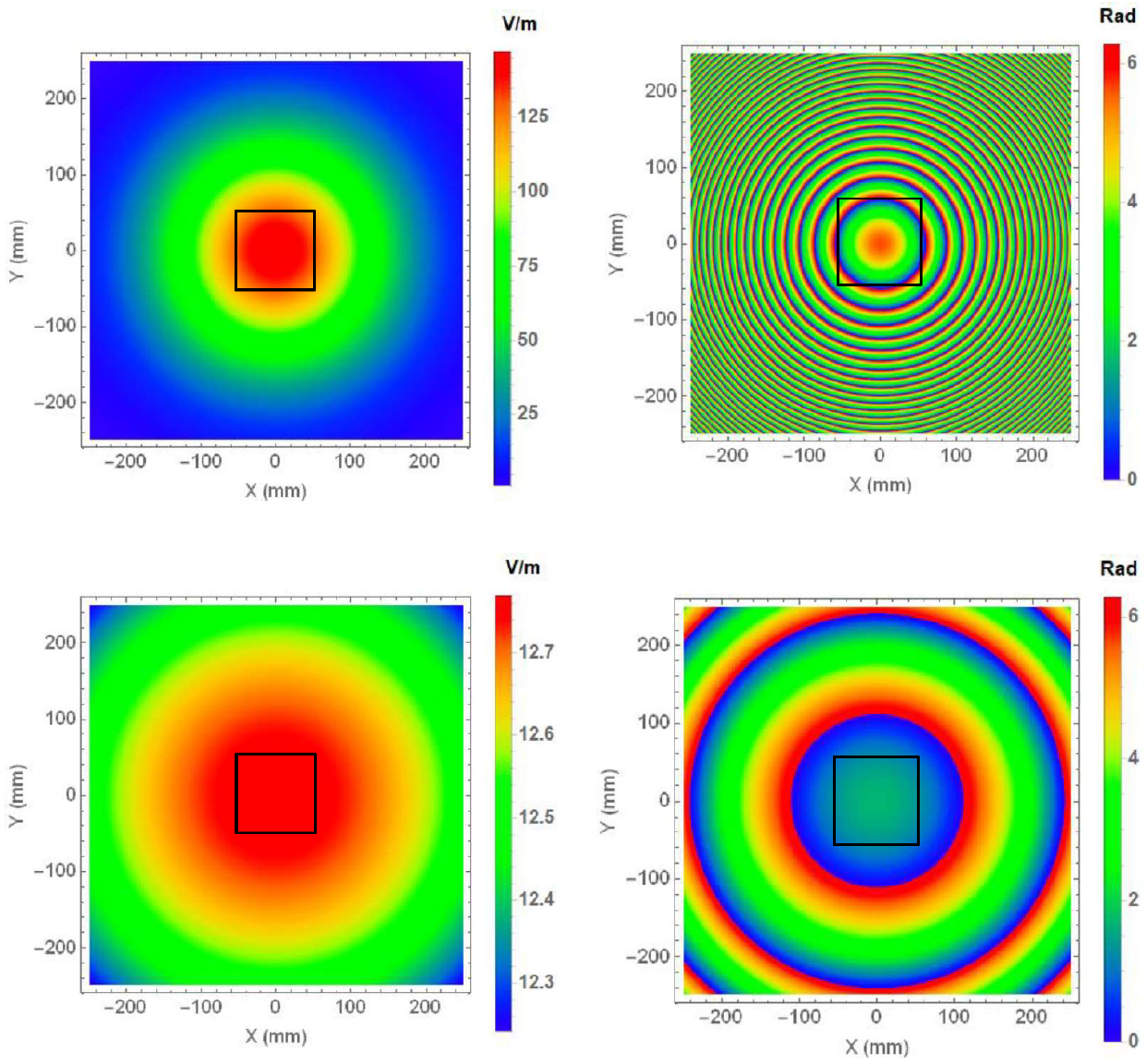}
  \caption{\new{Simulated beam pattern amplitude (left) and phase (right) from the source over the horn array of QUBIC
when the calibrator source is placed 1~m (top) and 10~m (bottom) from the aperture, with the position
and size of the horn array shown in black.}\label{fig:burke410}}
\end{figure}

Investigating the output of different horns in the array provides
another insight into the coupling to the incident source
radiation.  Incident radiation couples with the
horn and excites it, causing it to emit a beam which is approximately a $12.9^\circ$
Gaussian beam in the far-field. For an on-axis source in
the far-field of the instrument, the radiation over the array is
uniform and the coupling the same for all of the horns,
meaning that all the horns will emit a Gaussian-like beam of the same
amplitude and the same phase.  If the incident radiation
is not uniform over the array, as is the case for the source at 1~m,
then it will not couple to the horns in the same way and the horn
output will be different.
%% , as seen in figure~\ref{fig:burke411}.
%% \begin{figure}[t]
%%   \centering
%%   \includegraphics[width=0.95\linewidth]{burke_fig4_11}
%%   \caption{\new{Simulated beam profiles (intensity) for two different horns in
%%       the array when the source is at 1~m distance (top), 5~m (middle)
%%       and 10~m (bottom) calculated at 60~cm from the horn exit
%%       apertures.}\label{fig:burke411}}
%% \end{figure}
When the source is farther away, the coupling across the array becomes
more uniform.  All of the horn beams have the same profile.  The
difference in coupling can be seen as the difference in the peak
intensity.  With the calibration source at 10~m distance, the beams
from the two sample horns are \nnnew{similar}.

The PSF calculated for \nnnnew{some of} the calibrator source
distances investigated are shown in
figure~\ref{fig:burke412}~\cite{BurkePhD}. With the source placed 1~m
from the array, the image on the focal plane \nnnew{has more relative
  power spread throughout the image and a less clearly visible
  cross-like structure compared to the image with the source further
  away.}  A source distance of 10~m shows a pattern that
resembles the expected PSF figure~\ref{fig:burke412} (middle). This
implies that the source distance of 11.4~m is sufficiently faraway for
the calibrator to be used to measure the PSF of the TD. An optical
path of 11.4~m was achieved by placing the source on a lab wall and
using a mirror to reflect it into the QUBIC-TD aperture.
\begin{figure}[t]
  \centering
  \includegraphics[width=0.95\linewidth]{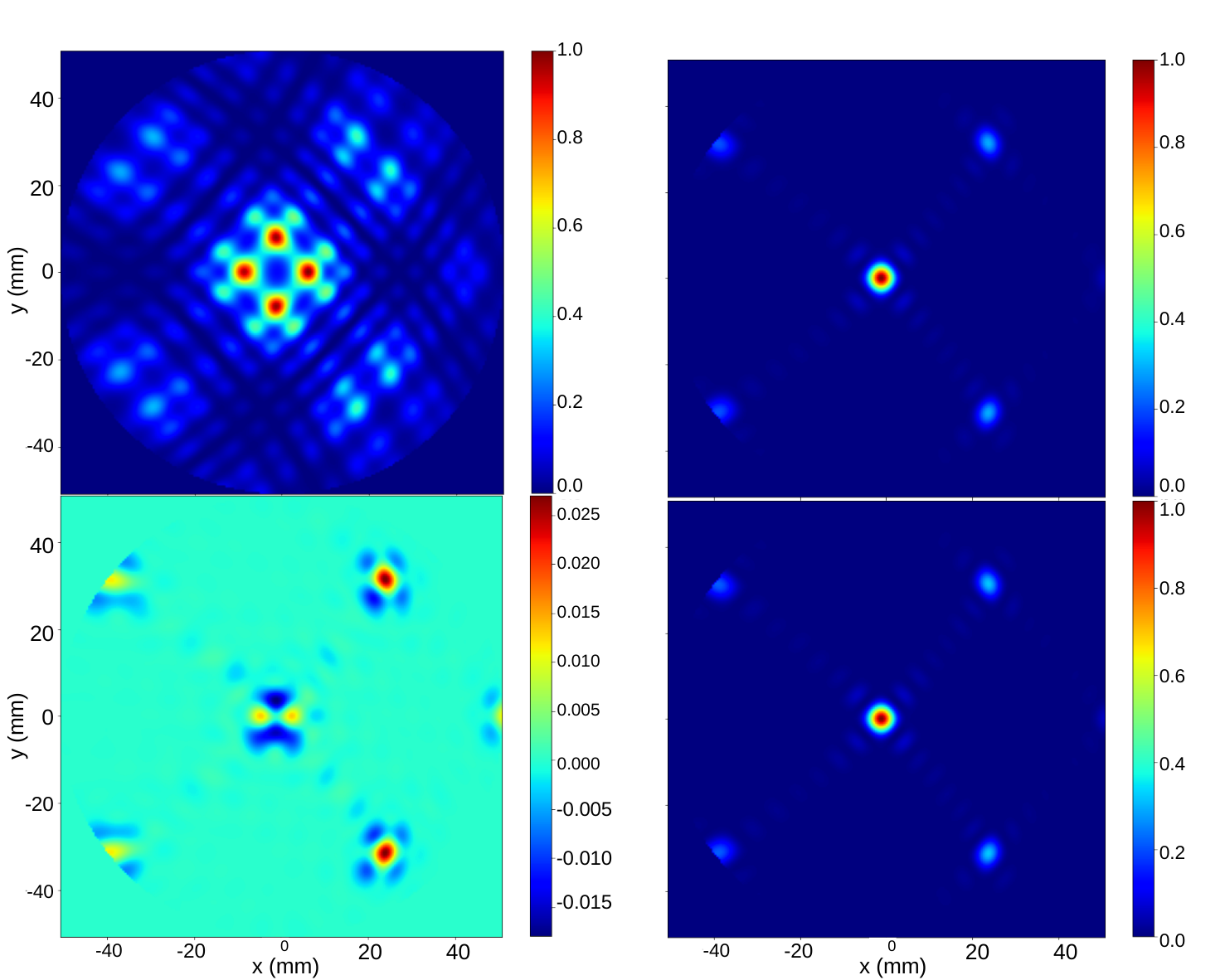}
  \caption{Simulated intensity of the interference pattern on the
    QUBIC focal plane when observing the calibration source at various
    distances from the instrument aperture.  Clockwise from top-left:
    source at 1~m, source at 11.4~m, source at infinite distance,
    \nnnnnew{difference between the source at 11.4~m and at infinity.  The
    colour bar gives the normalized intensity at each distance.  The
    peak residual is 0.025.}
    %% A diagonal cross-cut the patterns at 11.4~m (blue) and infinite
    %% distance (red) is shown (middle panel) along with the predicted
    %% PSF (blue) for comparison.  The predicted PSF was calulated using
    %% a plane wave (point source at infinity) as input.  The residuals
    %% between the two are shown in the bottom panel.
    \label{fig:burke412}}
\end{figure}

%% {A cut of the patterns at 5~m (blue) and 10~m (red) is shown (bottom)
%%   along with the predicted PSF (black) for comparison.  The predicted
%%   PSF was calculated using a plane wave (point source at infinity) as
%%   input.}
}

\clearpage
\section{Measurement Results and Analysis}
\label{sec:results}
\subsection{Spectral Response}
\label{sec:spectral}
\begin{figure}[t]
 \centering
 \nnewfig{\includegraphics[width=0.9\linewidth]{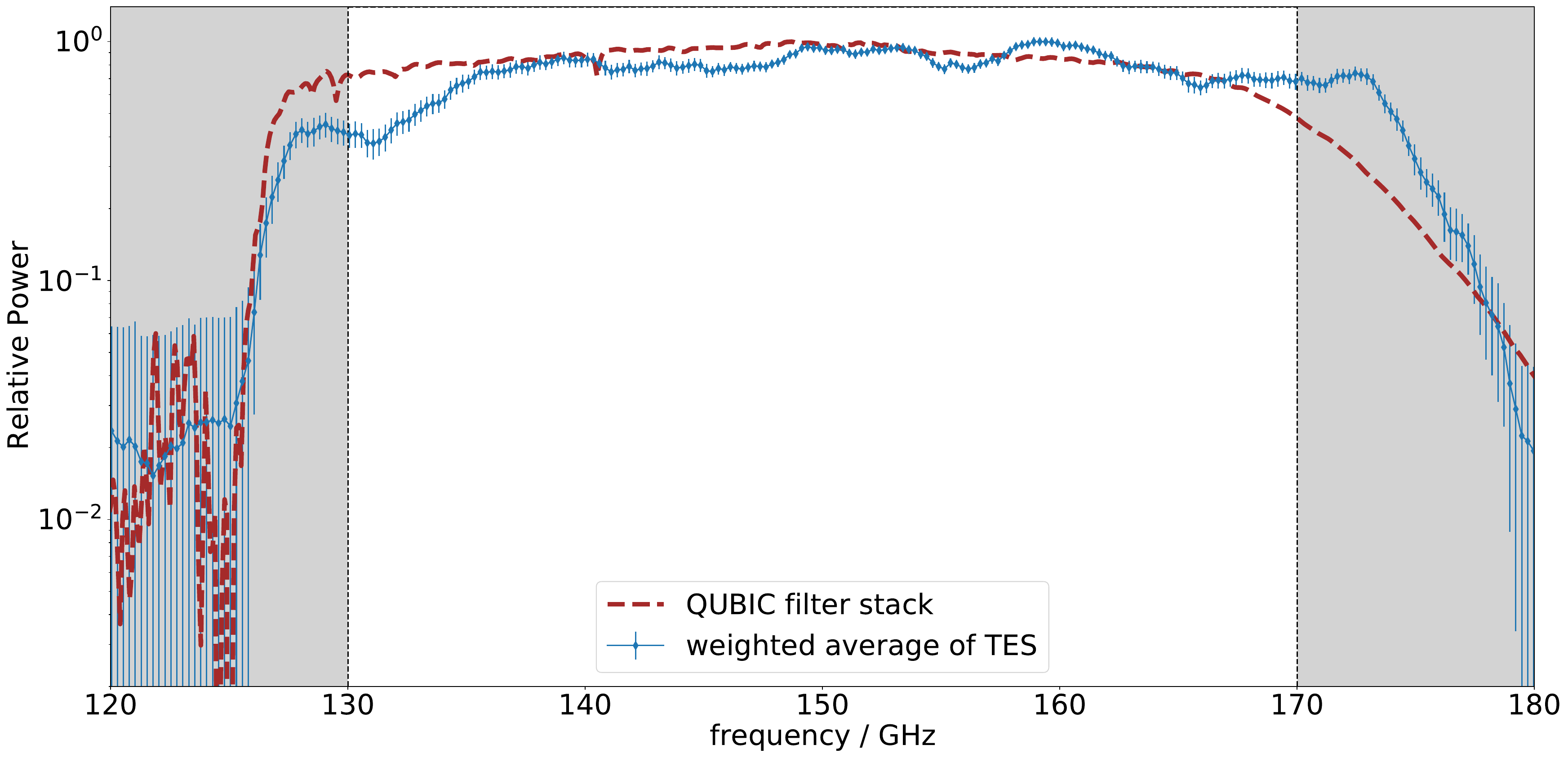}}
 \caption{Bandpass of the QUBIC-TD.  This was measured by stepping
   through the frequencies of the calibration source and measuring the
   relative power on the TES at each frequency.  \nnew{The expected
     profile from the combined effect of filters and the horn array is
     shown as the brown-dashed curve.  Note that the calibration
     source performance is uncertain outside the band 130~GHz to
     170~GHz (area shown in grey, and see main
     text).}\label{fig:bandpass}}
%% The calibration source output power is shown in red.  
%%  \nnew{corrected for spectral variations in output   power of the calibration source}
\end{figure}

The spectral response was measured using the calibration source (see
section~\ref{sec:calsource}).  The calibration source was modulated
with a 1~Hz sine wave modulation and was stepped through frequencies across
the band from 120~GHz to 180~GHz, \nnnew{taking 60~seconds of data at
  each frequency setting.}  The calibration source output power was
measured independently by a zero-bias detector at the output of the
amplitude multiplication chain, as described in
section~\ref{sec:calsource}.  Synchronized demodulation of the
calibration source output and the bolometer response provides a high
signal-to-noise measurement which also accounts for any spectral
variation in the calibration source output power.
Figure~\ref{fig:bandpass} shows the measured profile.  \nnnew{The
  error bars are statistical and calculated from the uncertainty in
  the demodulation of the calibration source signal.} \nnew{The
  spectral response is flat to within 1.5~dB inside the range 136~GHz
  to 173~GHz, and drops by $\sim2.5$~dB at the nominal band edge of
  130~GHz.}

%% \nnew{The calibration source is designed to operate only within the
%%   band 130~GHz to 170~GHz.  The performance outside this band was not
%%   characterized by Virginia Diodes Incorporated (VDI) who supplied the
%%   calibration source.  An estimate of the performance of the
%%   calibration source below 130~GHz is taken from a similar product,
%%   the AMC-610, also supplied by VDI.  At frequencies above 170~GHz,
%%   the output power of the calibration source is expected to drop
%%   significantly.  However, for the purpose of the plot in
%%   figure~\ref{fig:bandpass}, it is assumed that the power output
%%   remains constant at frequencies above 170~GHz.
%%   Figure~\ref{fig:bandpass} shows the measured bandpass corrected by
%%   the spectral variations of the calibration source, which is shown in
%%   red.}

%% proposed by Andrea Tartari (modified)
\nnew{ The calibration source is designed to operate between 130~and
  170~GHz (shown as the white region bordered by vertical dashed lines
  in figure~\ref{fig:bandpass}). Its use below and above these limits
  was not characterized by the supplier, Virginia Diodes Incorporated
  (VDI) and it should not be used to make quantitative conclusions due
  to uncertainty in the tuning frequency and in the level of the
  output power outside the designed spectral range.  Nevertheless, we
  operated the calibration source outside its nominal range, to obtain
  a qualitative behaviour of the spectral response of the intrument
  below and above the desired frequency range.  We also placed a layer
  of Eccosorb material AN72 at the mouth of the calibration source
  corrugated feedhorn in order to attenuate the high power output of
  the source.  This provides 25~dB of attenuation with a spectrally
  flat profile \nnnew{($\pm1$\% across the band)}, as measured by a vector network analyser at the
  laboratory of the University of Milan.}

%% We used the spectral
%%   shape of the source power output in the nominal band to deconvolve
%%   the intrinsinc instrument bandpass from the overall signal, as shown
%%   in figure~\ref{fig:bandpass}.  

%% The
%% in-band result is the effect of the stack of filters, at least down to
%% $\sim$ -6 dB.  Before the deployment of the instrument in the field we
%% plan to calibrate the out-of-band power supplied by the source for the
%% definitive frequency band reconstruction.
%% }

%% The plot shows the average of all operational TES (181 detectors, see
%% Piat et~al. \cite{2020.QUBIC.PAPER4}).

\setcounter{footnote}{0} The QUBIC-TD optical path includes
\nnnnew{13} multi-mesh blocking filters manufactured by QMC
Instruments Ltd~--~Thomas Keating Instruments
Ltd{\footnote{\href{http://www.terahertz.co.uk/}{http://www.terahertz.co.uk/}}}.
These are particularly important in the laboratory environment which
has a background thermal loading at room temperature, in contrast to
what will be the situation for the instrument when it is on the
observing site and pointing at the sky with a background thermal
loading on the order of 10~K.  The \nnnnew{13} filters are distributed
along the optical path inside the cryostat \nnnnew{and are heat sunk
  to} the different temperature stages \nnnnew{between 300~K and
  0.3~K.}  For more details on the optical design see O'Sullivan
et~al.~\cite{2020.QUBIC.PAPER8} and for the cryogenic design, see
Masi~et~al.~\cite{2020.QUBIC.PAPER5}.  In order to further attenuate
the 300~K environment in the laboratory, the QUBIC-TD has a neutral
density filter which is a thermally evaporated gold layer on a
$1.5\,\mu\mathrm{m}$ mylar substrate.  It was tested to give a 9\%
transmission at room temperature.  The combined filter profile is
shown as a brown-dashed curve in figure~\ref{fig:bandpass} and
includes all the filters in the optical train as well as the horn
waveguide cut-off profile.  The back-to-back horn array has a
waveguide ``throat'' section with a diameter that creates a
low-frequency cut-off at 125~GHz.  Details of the horn array are
described in Cavaliere et~al.~\cite{2020.QUBIC.PAPER7}.

\nnew{ The measured spectral profile differs from the expected profile
  at the band edges with additional suppression at the low frequency
  end, and additional transmission at the high frequency end.  The
  difference is $\sim2.5$~dB excess suppression at 130~GHz and
  $\sim4.5$~dB at 173~GHz.  The differences will be investigated in
  future measurements, however, this does not have a significant
  effect on the overall performance of the QUBIC-TD.  In particular,
  the measured bandpass falls well within the atmospheric window
  between the 119~GHz oxygen line and the 183~GHz water line.}

\afterpage{\clearpage}
\subsection{Half Wave Plate Polarization Rotation Test}
\label{sec:hwp}

%% A functionality test was carried out of the HWP rotator
%% mechanism while at the same time, the calibration source was operating.
%% Measurements were taken at each of the 7~evenly spaced positions of
%% the HWP from $0^\circ$ to $90^\circ$ (spacing of
%% $15^\circ$).

%% Figure~\ref{fig:hwp_signals} shows the signal
%% %, after Fourier filtering,
%% for \new{a TES near the centre of the focal plane} at different HWP
%% positions.  The peak-to-peak amplitude clearly varies with HWP
%% position.  Each position is shown in a different colour.  \new{The HWP
%%   orientation in position~\#3 rotates the calibration source
%%   polarization such that it is orthogonal to the polarizing grid (see
%%   figure~\ref{fig:qubicschematic} and
%%   section~\ref{sec:qubicdescription}).}

\begin{figure}[t]
  \centering
  \nnewfig{\includegraphics[width=0.99\linewidth]{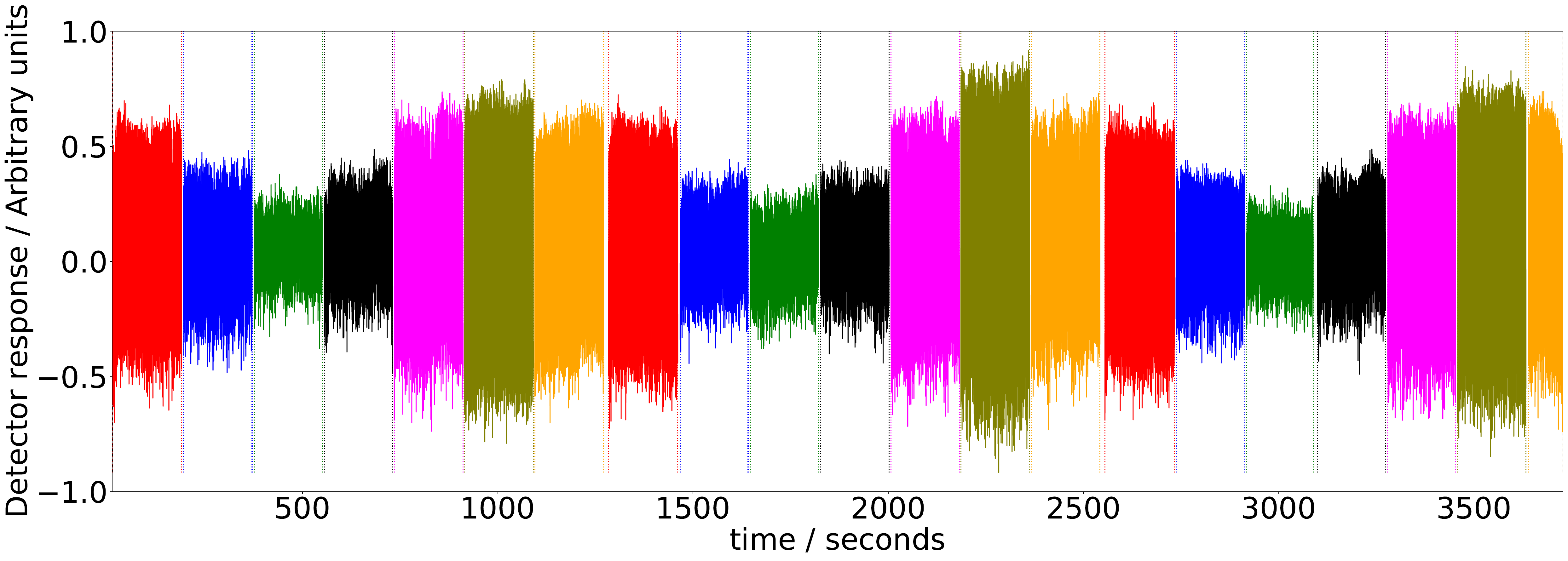}}
  \caption{\label{fig:hwp_signals}Peak-to-peak amplitude of the
    detected calibration source for different HWP positions as
    measured with \new{a TES near the centre of the focal plane.  The
      7~positions are spaced by $15^\circ$ in order to span
      $90^\circ$ between position~1 and position~7.}}
\end{figure}

\begin{figure}[t]
  \centering
  \nnewfig{
    \begin{minipage}[t]{\linewidth}
    \includegraphics[width=0.99\linewidth]{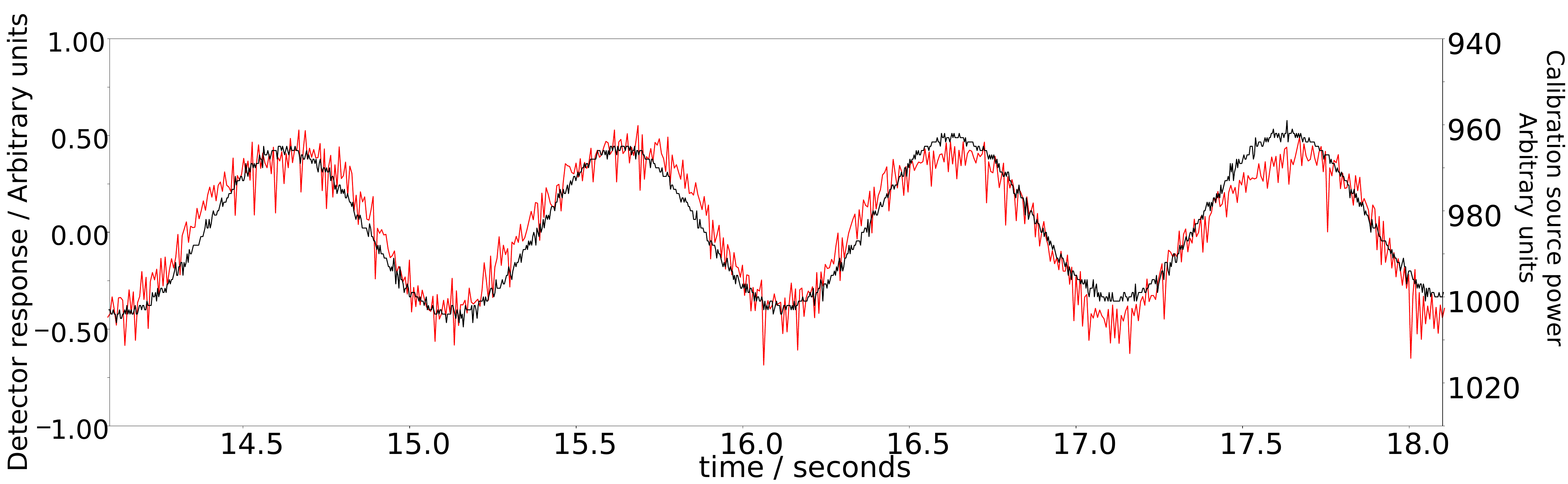}\\
    \includegraphics[width=0.99\linewidth]{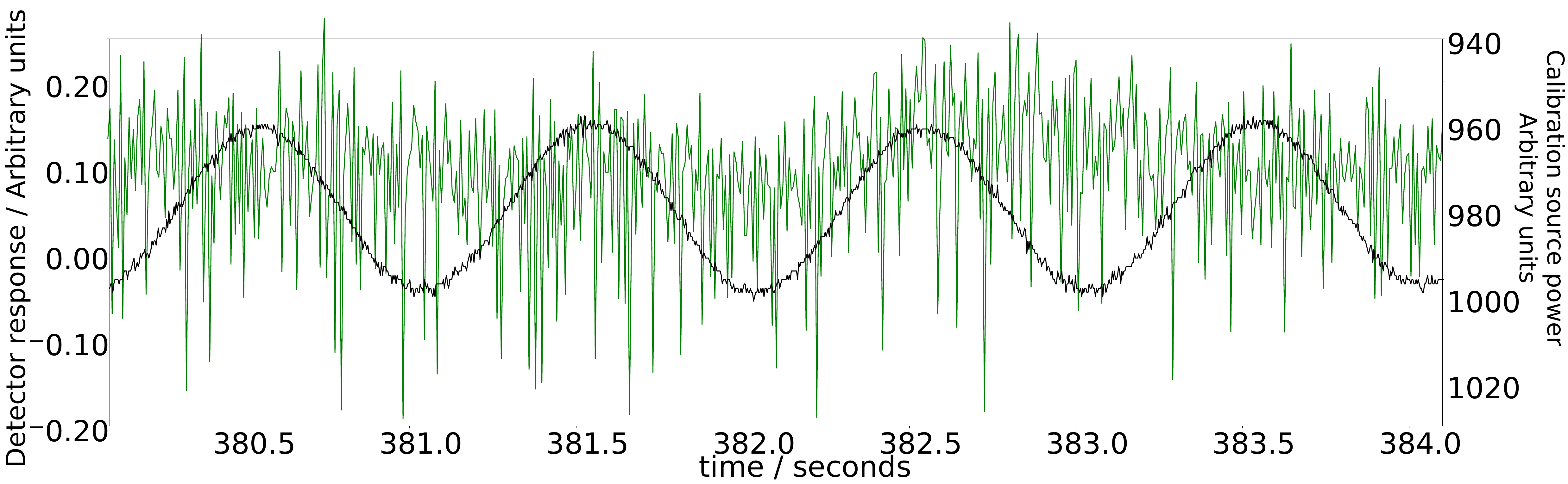}
    \end{minipage}
    }
  \caption{
    \label{fig:hwp_signals_zoom}
    Close-up view of the peak-to-peak amplitude of the detected
    calibration source for the HWP in \new{position~\#1~(top) and
      position~\#3~(bottom) as measured with a TES near the centre of
      the focal plane.  The black curve is the power of the
      calibration source as measured by the power monitor (see
      section~\ref{sec:calsource}).  It is plotted over the measured
      response of the bolometer (top, red curve for position~\#1 and
      bottom, green curve for position~\#3).  The HWP in position~\#3
      rotates the incoming calibration source polarization such that
      it is orthogonal to the polarizing grid.  This is seen in the low level
      of modulation in the signal.}}
\end{figure}

\begin{figure}[t]
  \centering
  \newfig{\includegraphics[width=0.85\linewidth]{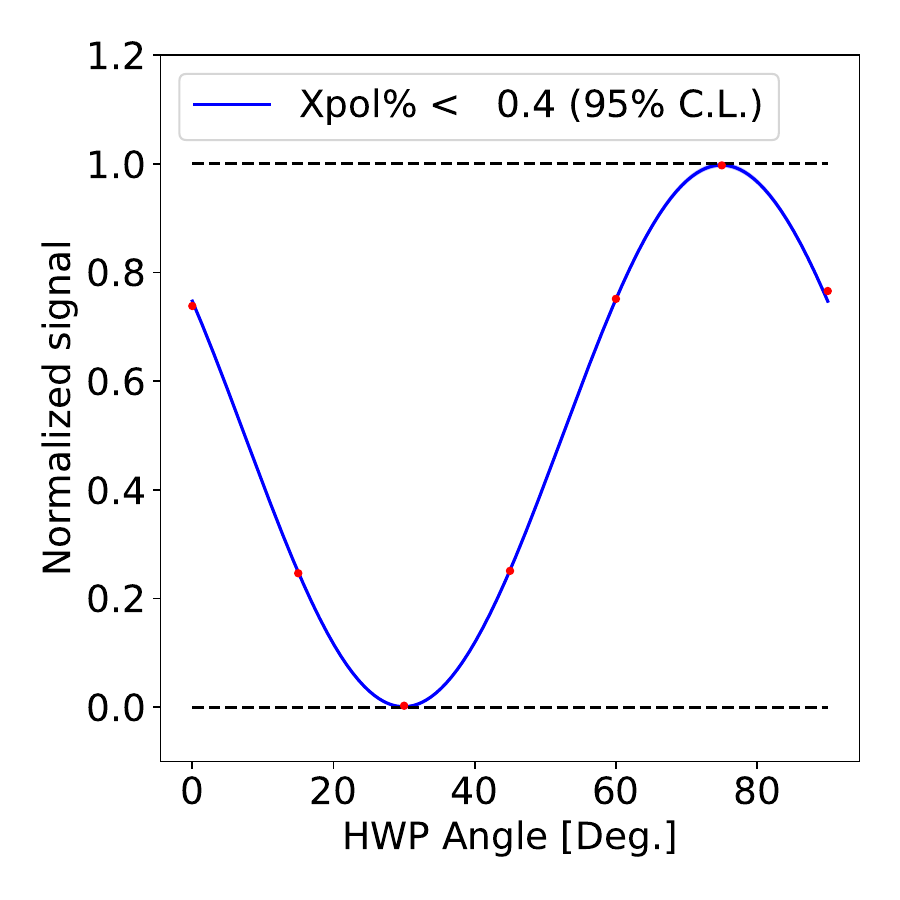}}
  \caption{\label{fig:hwp_amplitude}Amplitude of the detected
    calibration source at 150~GHz as measured with \new{a TES near the
      centre of the focal plane} for the HWP in the different
    positions and fitted to a sine curve.}
\end{figure}

A functionality test of the HWP rotator mechanism was carried out
while, at the same time, the calibration source was operating. The
calibration source, composed of a rectangular waveguide and a
corrugated horn, is linear polarized with cross-polarization
\nnnew{less than $-30$~dB, as quoted by the supplier of the corrugated horn, Custom
  Microwaves
  Inc.\footnote{\url{https://custommicrowave.com/}}}. Measurements
were taken at each of the 7~evenly spaced positions of the HWP from
$0^\circ$ to $90^\circ$ (spacing of $15^\circ$).
Figure~\ref{fig:hwp_signals} shows the signal for a TES near the
centre of the focal plane at different HWP orientations (shown in
different colours). The peak-to-peak amplitude clearly shows
repeatable variations with HWP position. The HWP orientation in
position~\#3 rotates the calibration source polarization such that it
is orthogonal to the polarizing grid (see
figure~\ref{fig:qubicschematic} and
section~\ref{sec:qubicdescription}).
Figure~\ref{fig:hwp_signals_zoom} shows a \nnnew{subset of 4~seconds}
of the calibration source signal measured by \new{a TES near the
  centre of the focal plane} with the HWP in position~\#1 \new{and in
  position~\#3}.  \nnnew{There are 180~seconds of
  data at each of the~7 HWP positions, giving high signal-to-noise and
  the small error bars seen in figure~\ref{fig:hwp_amplitude}.}

The amplitude at each position is plotted and fitted to a sine curve,
\new{with the amplitude normalized,} as shown in
figure~\ref{fig:hwp_amplitude}.  The signal in each case is measured
by the RMS of the TES data while the source was modulated.  The
measurement is the source modulation amplitude together with the RMS
of the noise, quadratically added. As a result the minimum value is
dominated by noise.  The maximum signal occurs when the HWP is between
position~6 and~7.  This measurement represents a direct measurement of
the cross-polarization response of the QUBIC-TD.
\clearpage
The cross-polarization contamination at 150~GHz \nnnew{for the TES
  near the centre of the focal plane} is compatible with zero to
within 0.4\% at 95\% confidence level.  \nnnew{We use the definition
  of cross-polarization described by eq.~1 of
  Ludwig~\cite{1973ITAP...21..116L}.}  \nnnew{The full analysis for all
  working detectors in the focal plane gives a median value of 0.61\%
  at 95\% confidence level~\cite{2020.QUBIC.PAPER6}.}
  
\begin{figure}[t]
  \centering
  \newfig{\includegraphics[width=0.8\linewidth]{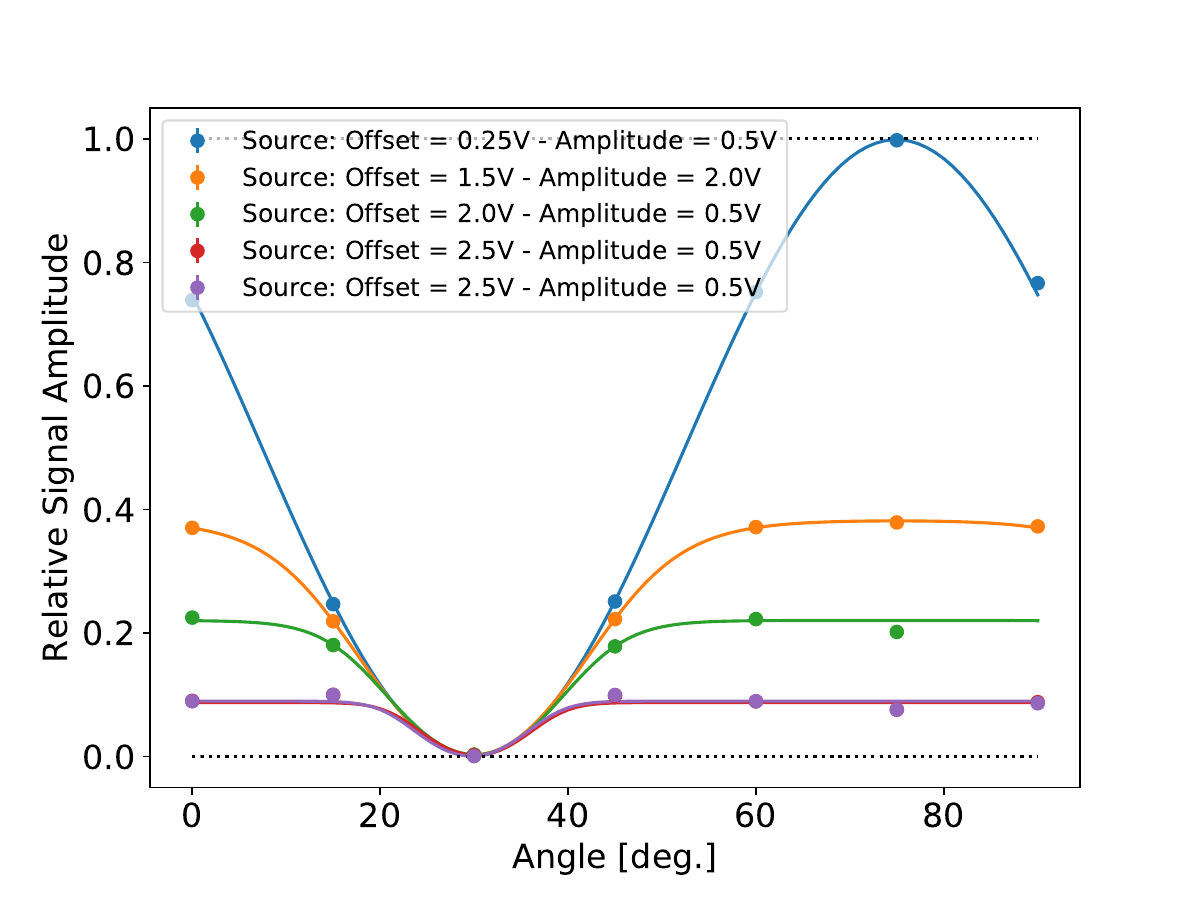}}
  \caption{\label{fig:hwp_saturation}Same as
    figure~\ref{fig:hwp_amplitude} but with the calibration source set
    at various power output levels.  The effect of saturation is
    evident for the source at high power.  The HWP angle
    of maximum signal damping is consistent between all the
    measurements.}
\end{figure}

\new{The calibration source can be modulated with a choice of
  modulation frequency, amplitude, and total power output (see
  section~\ref{sec:qubicdescription}).  The power offset and amplitude
  can lead to saturation of the TES detectors, as shown in
  figure~\ref{fig:hwp_saturation}.  The detector in this case is
  saturated at HWP positions which rotate the calibration source
  polarization closer to co-aligned with the polarizing grid.  The
  amplitude plotted in figure~\ref{fig:hwp_saturation} is the
  normalized amplitude such that the HWP rotation curve fits a sine
  curve with peak-to-peak amplitude of~1.  A sine curve fitting the
  lower part of the curve but not the upper part of the curve is an
  indication of saturation.  Figure~\ref{fig:hwp_saturation} is used
  to determine optimal calibration source settings so that the
  detector is not saturated at any HWP position.  \nnnew{Saturation is
    calculated using the saturation parameter as described in
    D'Alessandro et~al.~\cite{2020.QUBIC.PAPER6} (eq. 9.2).}  For the
  TES in figure~\ref{fig:hwp_amplitude} the saturation level is less
  than $0.23$\%.  }

As polarization is selected before the horn array (see section~\ref{sec:qubicdescription}), the
cross-polarization leakage is independent of the location in the focal
plane.  Therefore, the cross-polarization leakage is expected to be
the same for all TES.  The example shown in
Figures~\ref{fig:hwp_amplitude} and~\ref{fig:hwp_saturation} are for
\new{a TES near the centre of the focal plane} which has the best
signal-to-noise for this measurement due to the pointing.
%% \begin{figure}[t]
%%   \centering
%%   \includegraphics[width=0.8\linewidth]{hwp_phase}
%%   \caption{\label{fig:hwp_phase}Distribution of the trough in the
%%     amplitude for various source output power (see
%%     figure~\ref{fig:hwp_saturation}).}
%% \end{figure}
%% Figure~\ref{fig:hwp_phase} shows the cumulative
%% distribution of the measured cross-polarization for all TES with a
%% median of 2.3\%.  It is largely dominated by measurement uncertainty
%% as shown by the orange curve in figure~\ref{fig:hwp_cumulative}.
%% \begin{figure}[t]
%%   \centering
%%   \includegraphics[width=0.8\linewidth]{hwp_xpol_cumulative}
%%   \caption{\label{fig:hwp_cumulative}Cumulative distribution of the
%%     measured cross-polarization for all TES with a median of 2.3\% (blue curve).
%%     It is largely dominated by measurement uncertainty as shown by the
%%     cumulative distribution of the uncertainties (orange curve).}
%% \end{figure}

The cross-polarization contamination was measured with the
calibration source tuned to 150~GHz.  This is therefore a measurement
at a single, essentially monochromatic frequency.  Future test campaigns
will measure the cross-polarization contamination at frequencies
across the band in order to determine the integrated response of the
system for the full bandpass.

\afterpage{\clearpage}
\subsection{Self-calibration and the Measurement of Fringes}
\label{sec:selfcal}

Self-calibration is a technique developed for aperture synthesis in
radio interferometry.  This technique evolved from the original idea
in the 1970's of
``phase-closure''~\cite{1974A&AS...15..417H,1976A&A....50...19F,1978ApJ...223...25R}
to become in the early 1980's
``self-calibration''~\cite{1981MNRAS.196.1067C,1984ARA&A..22...97P}.
See Cornwell \& Fomalont~\cite{1999ASPC..180..187C} for a detailed overview.  Precise
knowledge of the calibration source is not required, as long as it is
a strong and stable point source.  The large number of baseline visibilities
allows us to solve for many unknowns, including the gain and phase
corrections \nnnew{of each detector}, without requiring knowledge of the source amplitude.

%\subsection{Fringes}
\label{sec:fringes}
In order to advance towards the full analysis of self-calibration, a
key component is the ability to measure fringes with QUBIC.  If
fringes can be measured with a horn pair, then the full analysis can
be done once the fringe measurement is done for all horn pairs.  By
measuring the fringe pattern of a single pair of horns we demonstrate
the feasibility of doing self-calibration with the bolometric
interferometer.

\begin{figure}[t]
 \centering
   \newfig{\includegraphics[height=0.2\textheight]{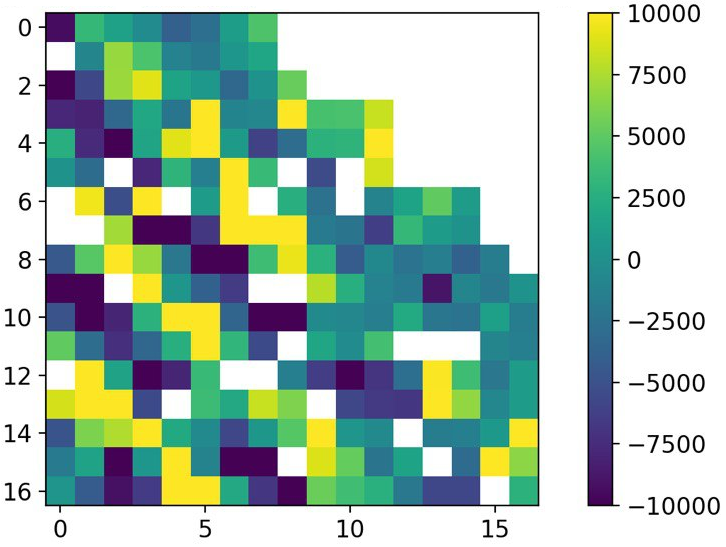}
     \includegraphics[height=0.2\textheight]{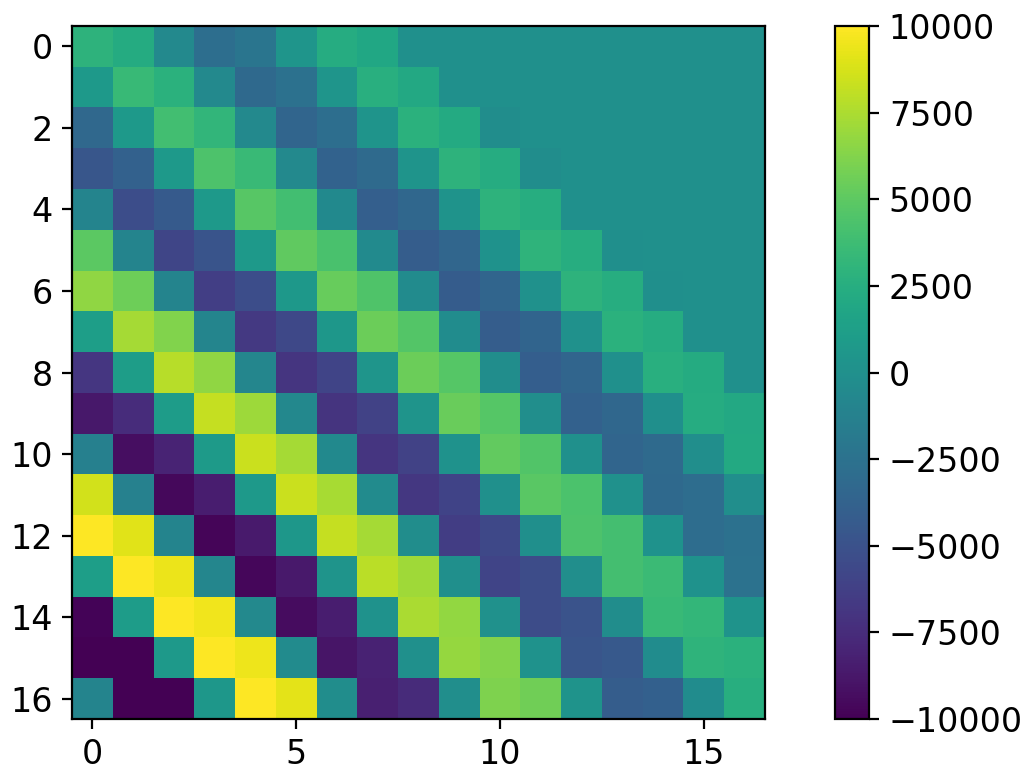}}
 \caption[Fringes]{Fringes on the QUBIC TES array.
   \textbf{left:}~~Measured fringes corresponding to the baseline
   between horns~25 and ~57.  The fringes are clearly visible as
   bright, diagonal lines across the detector array.
   \textbf{right:}~~Simulation of the expected fringe pattern for the
   same baseline pair generated by \swname{qubicsoft}. The measured
   fringe locations \nnnew{correspond} with the simulated image.  The
   amplitude of the fringe lines are expected to become fainter with
   distance from the centre of the focal plane.  This was not measured
   because of saturation of the detectors. \nnnnnew{Pixel intensities
     are the same arbitrary units scaled to the peak of the
     measurement for both images.}}
 \label{fig:fringes}
\end{figure}

The QUBIC-TD successfully measured fringes between a pair of horns
(figure~\ref{fig:fringes} left).  \new{The measured fringe pattern in
  figure~\ref{fig:fringes} (left) is the interference pattern of one
  baseline which corresponds to a set of two horns.}  It is the
derived image after analysing measurements of images with all horns
open, with two horns closed, with one horn closed, and with the other
horn closed.  \nnnew{This measurement takes approximately 30~minutes,
  cycling through the switch configurations (horn~`a' open, horn~`b'
  open; horn~`a' closed, horn~`b' open; horn~`a' closed, horn~`b'
  closed; horn~`a' open, horn~`b' closed).  Each configuration is
  measured for 20~seconds, and the timeline is co-added (folded by the
  cycle period).}  The result is the equivalent of having all horns
closed except the two of interest, as was shown in Bigot-Sazy
et~al~\cite{2013A&A...550A..59B}.  \new{It is necessary to make the
  measurement in this way because of the limitation required on heat
  dissipation.  Only a maximum of two switches can be activated
  (closed) at a time, as described in
  section~\ref{sec:qubicdescription}~(and see Cavaliere et~al.
  \cite{2020.QUBIC.PAPER7} for full details).  As a result, it is not
  possible to close all horn switches except for the two of
  interest. However, using the analysis steps described above,
  figure~\ref{fig:fringes} (left) is the result of interference
  between a single pair of horns.}

The fringes are expected to be fainter in proportion to the distance
to the centre of the focal plane (figure~\ref{fig:fringes} right).
This is not the case here because of saturation of the TES detectors.
As a result, the fringe amplitude appears to be relatively constant,
or near zero where saturated detectors were subtracted from one
another.  The problem with saturation \new{is mainly related to the
  high power of the calibration source.}  The calibration source can be attenuated by its
  bias voltage, which is used for signal modulation (see, for example,
  figure~\ref{fig:hwp_saturation} in section~\ref{sec:hwp}).  It can
  also be attenuated by moving the HWP, as seen in
  figure~\ref{fig:hwp_amplitude} of section~\ref{sec:hwp}.  Future tests will
\new{explore methods to attenuate the calibration source while still
  having enough signal-to-noise to measure the secondary lobes of the
  synthesized beam.}

\nnnew{ It is worth noting that the difficulty with saturation of TES
  is more problematic for the analysis of fringes than for the
  analysis of the HWP presented in section~\ref{sec:hwp}.  This is
  because the HWP measurements were done with a single TES at a time.
  This allows us to test the HWP performance with optimal settings on
  a given TES, and thus separate the HWP performance from the TES
  performance.
The fringe measurement requires constructing an image on the focal
plane, and this necessitates the use of multiple TES.  The TES
manufacturing is not perfectly homogenous (see Piat et~al.~\cite{2020.QUBIC.PAPER4} for details)
and the optimum bias settings vary somewhat from TES to TES.  The
electronics supplying the bias settings for the TES can only supply
the same bias to all 128 TES configured by each ASIC.  As a result,
some of the TES are not optimally biassed, and saturation could be a
problem for some while not for others.  A solution to the problem is
to make a sky map of the fringes, in the same way we made the map of
the synthesized beam.  In order to do this, a map must be made for
each horn-switch configuration (single horns open/closed and pairs
open/closed).  This will be time consuming, but it is done with the
artificial calibration source, and so can be done during times when
QUBIC is not measuring the sky.  Eventually, we will build a data base
of maps with all the horn configurations.
  }

\setcounter{footnote}{-1}
%% An initial analysis using the fringe measurement is carried out in
%% section~\ref{sec:hornorientation}.  The measurement is used to verify
%% the orientation of the horn array.

%\input{horn_orientation}

\afterpage{\clearpage}
\subsection{Synthesized full beam reconstruction}
\label{sec:syntheticbeam}

The QUBIC-TD synthesized beam maps were measured at five frequencies in
the range 130~GHz to 170~GHz by tuning the VDI calibration source (see
section~\ref{sec:calsource}) in steps of 10~GHz from 130~GHz.  For each
frequency measurement the calibration source was modulated at a period
of 1~second with a sinusoidal profile, and the TES signal was
demodulated in post processing.  QUBIC-TD was configured to point at an
elevation angle of 35$^{\circ}$ and to scan across azimuth at a
constant rate from $-25^\circ$ to $+25^\circ$.  The elevation angle
was then increased by $0.2^\circ$ and a new azimuth scan at constant
rate was done in the reverse direction from $+25^\circ$ to
$-25^\circ$.  Each azimuth scan takes nearly 8~minutes, and an entire
beam map at a given frequency is completed in 22~hours and 30~minutes.
This matches well with the cryogenic hold time of the cryostat.  The
entire beam map measurement was preprogrammed in a script executed by
the QUBIC instrument control software \swname{QubicStudio}.

\begin{figure}[t]
  \centering
  \newfig{\includegraphics[width=0.9\linewidth]{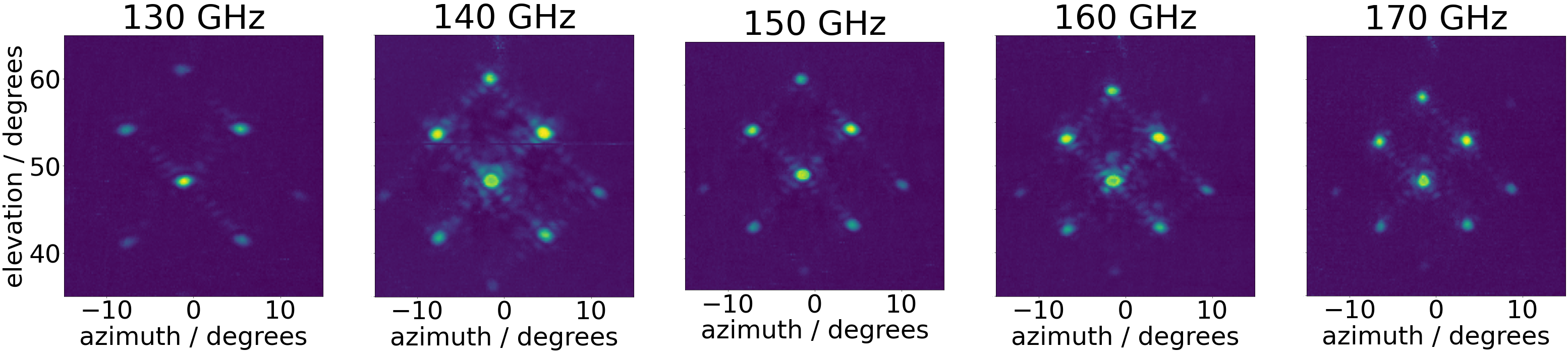}}
  \caption{\label{fig:synthbeam}Measured maps of the synthesized beam for
    \new{one TES detector}.  The multilobe synthesized beam is shown
    here for 130~GHz, \new{140~GHz,} 150~GHz, \new{160~GHz,} and 170~GHz.  An animated version of
    this plot with comparison to simulation can be viewed online at
    \href{https://box.in2p3.fr/index.php/s/bzPYfmtjQW4wCGj}{https://box.in2p3.fr/index.php/s/bzPYfmtjQW4wCGj}}
\end{figure}

\begin{figure}[t]
  \nnewfig{
    \begin{minipage}{\linewidth}
      \centering
      \includegraphics[width=0.6\linewidth]{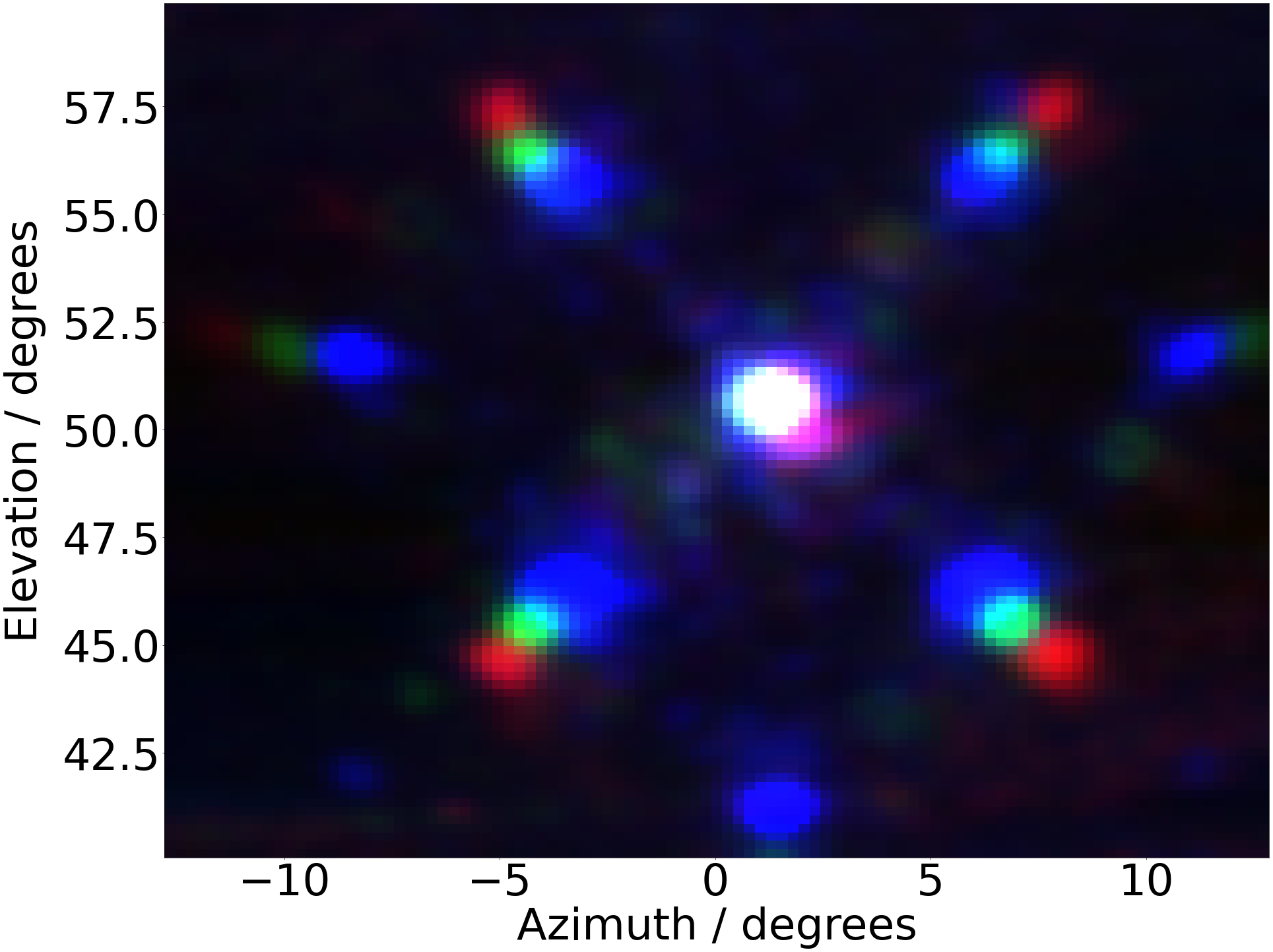}\\
      \includegraphics[width=0.95\linewidth]{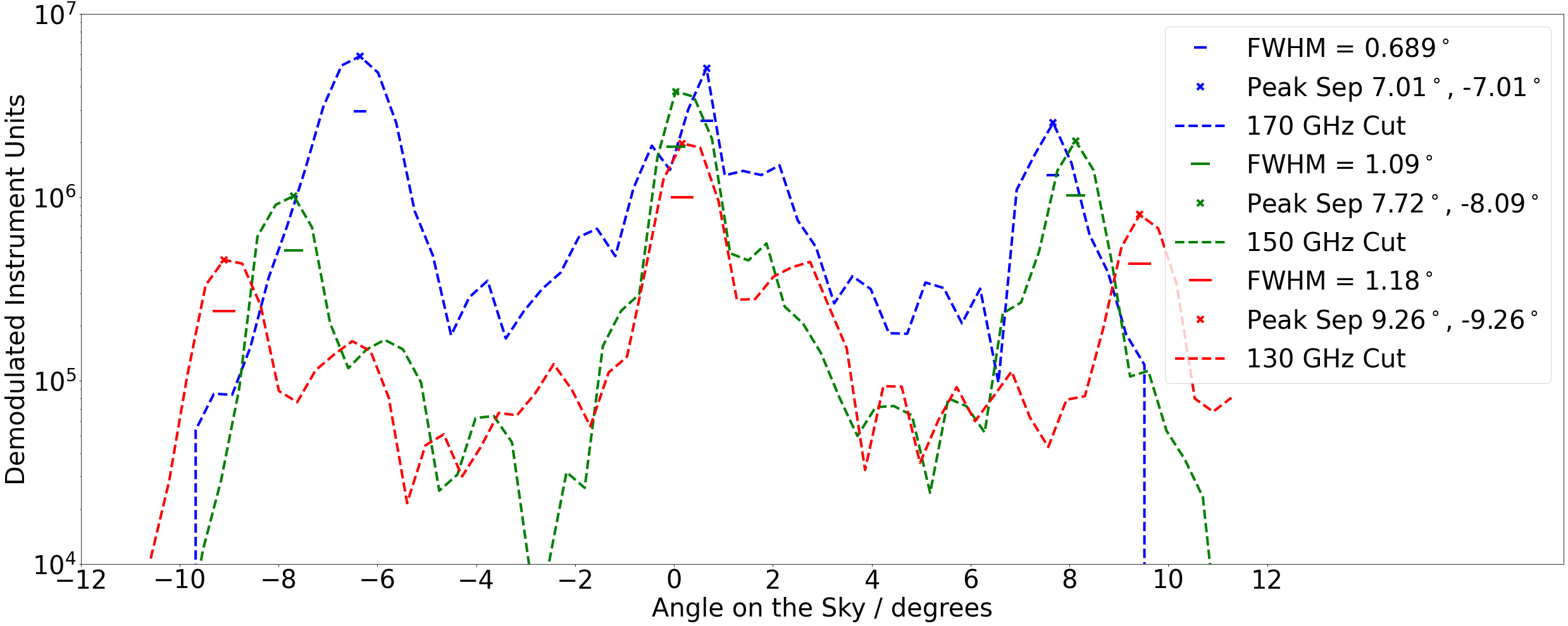}
    \end{minipage}
  }
  \caption{\label{fig:synthbeam_cuts}\new{Multichroic representation
      of the maps of figure~\ref{fig:synthbeam}.  Each frequency is
      assigned a colour, \nnew{130~GHz}~(red), \nnew{150~GHz}~(green),
      and \nnew{170~GHz}~(blue).  The plot above shows the
      coincident central lobe for each frequency and the secondary
      lobes which are closer to centre as frequency increases.  The
      plot below is the cross-cut along the diagonal from
      bottom-left to top-right of the image on the left.  The peak at
      170~GHz is truncated because of detector saturation.}  }
\end{figure}

The result of the beam mapping measurement campaign is a series of
maps for each TES pixel in the detector array.  This is 244~maps for
each frequency for a total of 1220~maps.  Figure~\ref{fig:synthbeam}
shows example maps for \new{one TES detector} at \new{five} frequencies.
The main lobe and secondary lobes are clearly visible and \new{are at the expected locations.}
\new{Figure~\ref{fig:synthbeam_cuts} shows the
  maps of figure~\ref{fig:synthbeam} in a single representation with
  three frequencies, 130~GHz, 150~GHz, and 170~GHz, assigned to colours red, green, blue.}

The secondary lobe locations depend on the frequency \new{of the
  incident radiation} while the main lobe is always at the same place.
Figure~\ref{fig:synthbeam_cuts} shows a cut across the beam maps of
figure~\ref{fig:synthbeam}.  The secondary lobes are closer to centre
\new{for higher frequencies}, as expected.  This result is the first
\nnew{hardware demonstration} of the capability of a bolometric
interferometer to operate as a spectral imager~(see Mousset
et~al. \cite{2020.QUBIC.PAPER2} for a detailed analysis).

\subsection{Map Making with Measured Synthesized Beams}
\label{sec:mapmaking}

Using the relative location and amplitude of all the peaks in the
synthesized beam for each of the TES, we can now project the data onto
the sky using optimal map-making to deconvolve from the effect of the
multiple peaks. In addition, by exploiting the spatial separation of
secondary peaks and the dependence of their location on the frequency
of radiation, the analysis of the TOD leads to spectral imaging as
described in detail in Mousset et~al.~\cite{2020.QUBIC.PAPER2}.
Spectral discrimination is necessary to disentangle foreground
contamination.  When observing the sky, this will result in an
unbiased CMB map as is shown using simulations in \nnew{Mousset
  et~al.}~\cite{2020.QUBIC.PAPER2}.

\begin{figure}[t]
  \centering
  \newfig{
    \begin{minipage}[h]{\linewidth}
      \centering
    \includegraphics[width=0.18\linewidth]{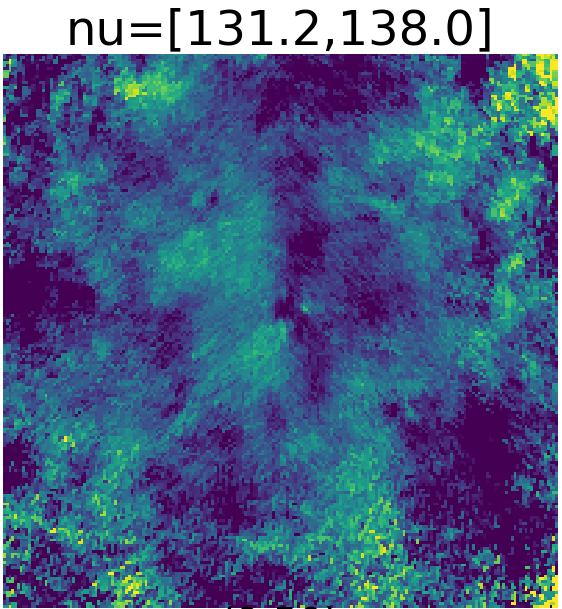}
    \includegraphics[width=0.18\linewidth]{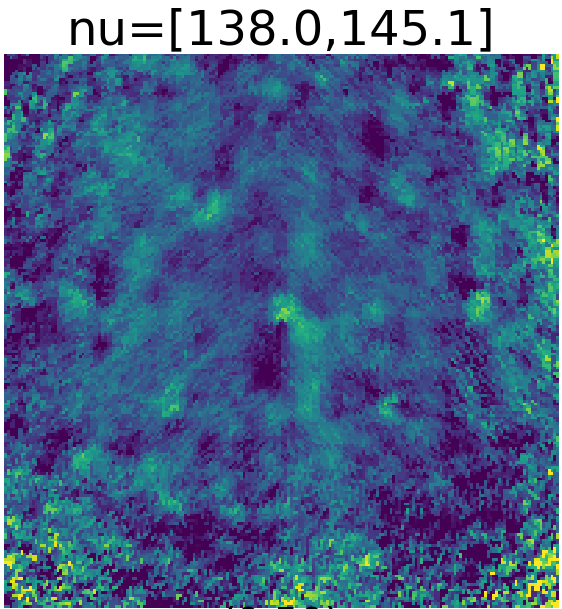}
    \includegraphics[width=0.18\linewidth]{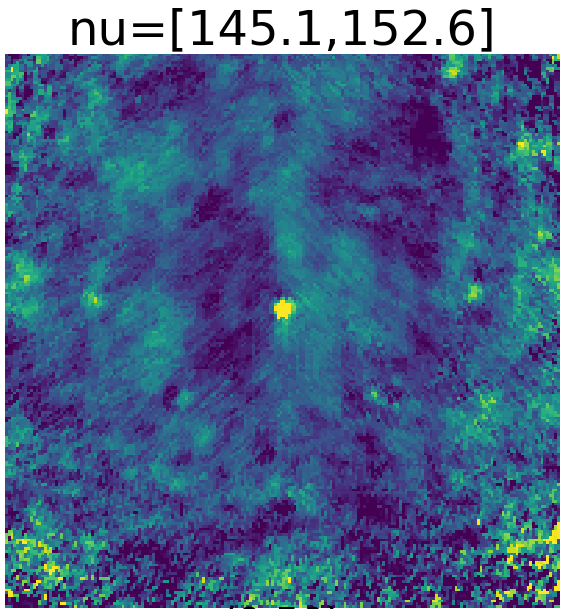}
    \includegraphics[width=0.18\linewidth]{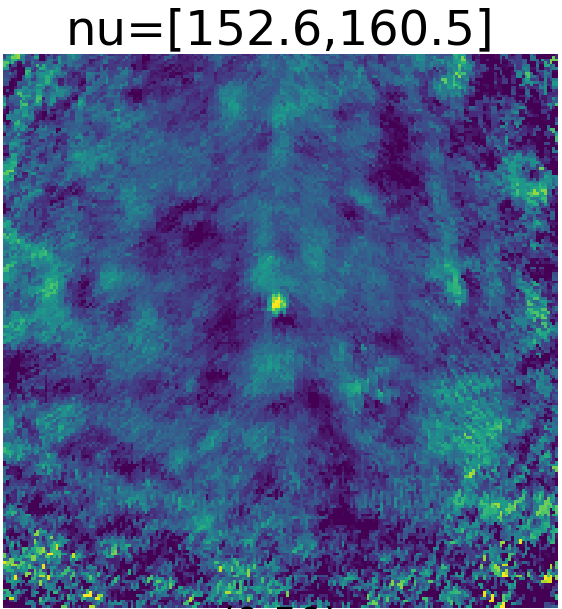}
    \includegraphics[width=0.18\linewidth]{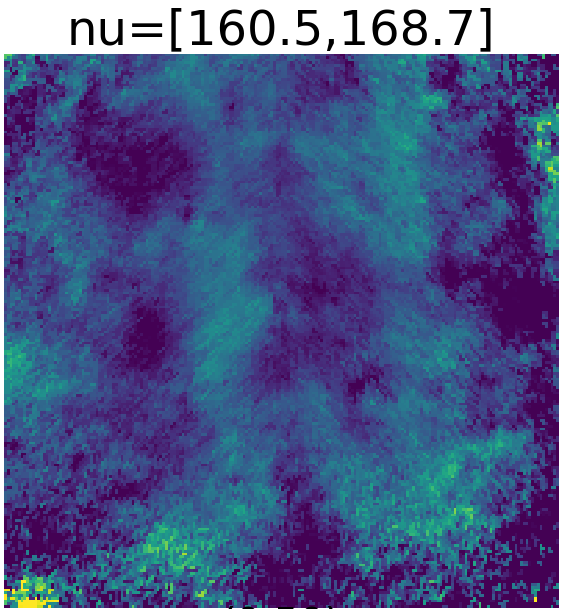}\\
    \includegraphics[width=0.7\linewidth]{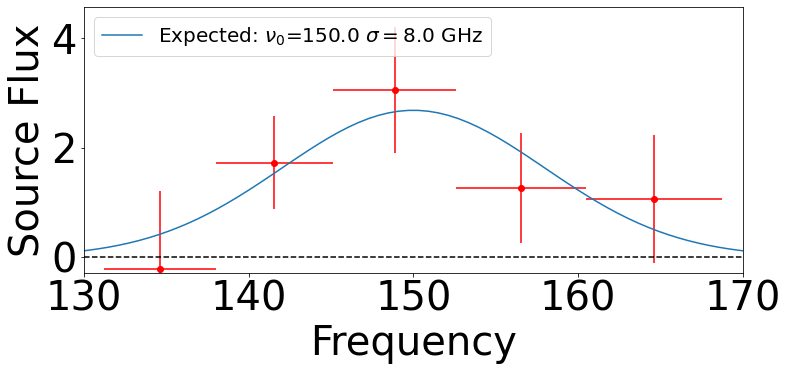}
    \end{minipage}
    }
  \caption{\label{fig:mapmaking}Calibration data with the source at
    150~GHz projected on the sky using our map-making software to
    deconvolve from the multiple peaked synthesized beam \new{and
      separate the physical band of the instrument into 5~sub-bands.
      The spectrum of the central pixel, in arbitrary flux units, is
      shown at bottom and matches the expected performance of spectral
      imaging by analysis of the TOD and applying the frequency
      dependence of the QUBIC beam as described in
      \nnew{Mousset et~al.}~\cite{2020.QUBIC.PAPER2}.}  }
\end{figure}

We have performed this exercise with the calibration data in order to
obtain an image of the point-like artificial calibration source
projected onto a sky-map.  \nnnnew{ The calibration source was set to
  emit at 150~GHz and using the alt-azimuth mount, \qtd\ was scanned
  across the source in azimuth and elevation.  TOD are obtained for
  each bolometer. We then apply our spectral imaging map-making
  algorithm with five sub-bands to a selection of 26~bolometers that
  do not exhibit saturation. The synthesized beam for each bolometer
  is realistically modeled in our map-making through a series of
  Gaussian profiles whose amplitudes, widths and locations are fit
  from a measured map of the synthesized beam for each bolometer (see
  figure~\ref{fig:synthbeam} for an example of the maps from one
  bolometer). The frequency evolution of this synthesized beam assumes
  linear scaling with wavelength. We were able to reconstruct a map of
  the point-like artificial calibration source as well as its location
  in frequency space.  The resulting image is shown in
  figure~\ref{fig:mapmaking}.  The reconstruction onto 5~sub-bands
  exhibits the expected point-like shape in the central frequency
  sub-band containing the emission frequency of the source at
  150~GHz.}  The calibration source, which was tuned to 150~GHz, is
fainter in adjacent bands, and not visible in the furthest bands.
\nnnnew{Our measurement matches the spectral resolution predicted by
  the Rayleigh criterion (see Mousset et~al.~\cite{2020.QUBIC.PAPER2},
  section 3.2 equation 3.1) shown as the blue line in
  figure~\ref{fig:mapmaking} (bottom).}

The successful mapmaking with the measured synthesized beam is
effectively an end-to-end checkout of the entire QUBIC system.  In
order for this exercise to be possible, all subsystems, interfaces
between subsystems, and all associated software must be functioning
correctly.  This includes scientific and housekeeping data
acquisition, telescope pointing control, and control and
synchronisation of all subsystems.  The software needed to make this
measurement includes the system control software, the data acquisition
software, data archiving and reading, and finally data analysis
software together with comparison to system simulation software.  The
resulting map of figure~\ref{fig:mapmaking} shows that all subsystems
are functioning correctly, and all subsystems are correctly managed
and synchronized together into the overall system.

\afterpage{\clearpage}
\section{Conclusion}
\label{sec:conclusion}
QUBIC-TD underwent an extensive campaign of testing in the laboratory at APC
in Paris.  Using an artificial millimetre-wave source in the
telescope far-field, \new{functional tests were carried out and
  several performance parameters were successfully measured.  Our
  measurement of the frequency-dependence of the synthesized beam is
  in excellent agreement with the theoretical prediction. This result,
  for the first time, opens up the possibility to use bolometric
  interferometry to perform spectral imaging. Based on these results,
  which were carried out on a single calibrating point source,
  preliminary simulations of QUBIC spectral imaging in presence of
  more realistic diffuse signals are very
  encouraging~\nnew{(Mousset~et~al.~\cite{2020.QUBIC.PAPER2})}.}  The
spectral response \nnew{has} the \nnew{required} bandpass
\nnew{profile for operation in the 130~GHz to 170~GHz band.}  The
polarization performance has less than 0.4\% cross-polarization
contamination at 150~GHz.  These measurements confirm that bolometric interferometry
is a viable method for the measure of CMB B-mode polarization.  In
particular, \new{the \nnew{measured} polarization performance makes us
  confident that QUBIC will be one of the most competitive experiments
  in terms of polarization purity.}

%% The emphasis on QUBIC design has been on the suppression and control
%% of systematic effects.  QUBIC employs bolometric interferometry and a
%% conservative polarization optical design which together leads to
%% control of \new{systematic effects} while still giving the
%% high sensitivity inherent in the use of wide band bolometers.

Deployment of QUBIC-TD on the scientific site of Alto Chorillo at 5000~m
altitude in Argentina is expected \new{by the end of 2022 for in-situ 
commissioning and preliminary data taking, as a precursor of the
QUBIC full instrument.}  QUBIC
will provide a clean polarization map of the sky and
together with \new{spectral imaging}, will have excellent separation
of foreground sources.

\acknowledgments

QUBIC is funded by the following agencies. France: ANR (Agence Nationale de la
Recherche) 2012 and 2014, DIM-ACAV (Domaine d’Intérêt Majeur-Astronomie et Conditions d’Apparition de la Vie), CNRS/IN2P3 (Centre national de la recherche scientifique/Institut national de physique nucléaire et de physique des particules), CNRS/INSU (Centre national de la recherche scientifique/Institut national et al de sciences de l’univers). Italy: CNR/PNRA (Consiglio Nazionale delle Ricerche/Programma Nazionale Ricerche in
Antartide) until 2016, INFN (Istituto Nazionale di Fisica Nucleare) since 2017.  Argentina: MINCyT (Ministerio de Ciencia, Tecnología e Innovación), CNEA (Comisión Nacional de Energía Atómica), CONICET (Consejo Nacional de Investigaciones Científicas y Técnicas).
 
D. Burke and J.D. Murphy acknowledge funding from the Irish Research Council under the Government of Ireland Postgraduate Scholarship Scheme.  D. Gayer and S. Scully acknowledge funding from the National University of Ireland, Maynooth. D. Bennett acknowledges funding from Science Foundation Ireland.

\afterpage{\clearpage}
\bibliographystyle{ieeetr}
\bibliography{qubic}  

\end{document}